\RequirePackage{lineno}
\documentclass[aps,titlepage,12pt,superscriptaddress]{revtex4}
\usepackage{epsfig}
\usepackage{float}
\usepackage{graphicx}
\usepackage{graphics}
\usepackage{times}
\usepackage{color}
\usepackage{multirow}
\usepackage{lettrine}
\usepackage{epstopdf}
\usepackage{amsmath}

\begin{document}

\title{Early exclusion leads to cyclical cooperation in repeated group interactions}

\author{Linjie Liu}
\affiliation{College of Science, Northwest A \& F University, Yangling 712100, China}
\affiliation{School of Mathematical Sciences, University of
Electronic Science and Technology of China, Chengdu 611731, China}

\author{Zhilong Xiao}
\affiliation{School of Mathematical Sciences, University of
Electronic Science and Technology of China, Chengdu 611731, China}

\author{Xiaojie Chen}
\email{xiaojiechen@uestc.edu.cn}
\affiliation{School of Mathematical Sciences, University of
Electronic Science and Technology of China, Chengdu 611731, China}

\author{Attila Szolnoki}
\affiliation{Institute of Technical Physics and Materials Science, Centre for Energy Research, PO Box 49, H-1525, Budapest, Hungary}

\begin{abstract}\noindent
\\
Explaining the emergence and maintenance of cooperation among selfish individuals from an evolutionary perspective remains a grand challenge in biology, economy, and social sciences. Social exclusion is believed to be an answer to this conundrum. However, previously related studies often assume one-shot interactions and ignore how free-riding is identified, which seem to be too idealistic. In this work, we consider repeated interactions where excluders need to pay a monitoring cost to identify free-riders for exclusion and free-riders cannot participate in the following possible game interactions once they are identified and excluded by excluders in the repeated interaction process. We reveal that the introduction of such exclusion can prevent the breakdown of cooperation in repeated group interactions. In particular, we demonstrate that an evolutionary oscillation among cooperators, defectors, and excluders can appear in infinitely large populations when early exclusion is implemented. In addition, we find that the population spends most of the time in states where cooperators dominate for early exclusion when stochastic mutation-selection is considered in finite populations. Our results highlight that early exclusion is successful in solving the mentioned enigma of cooperation in repeated group interactions.
\end{abstract}


\maketitle

\noindent

\vbox{}
\noindent\textbf{1. Introduction}

\noindent Cooperative behaviour provides generous benefits for group members, but it is vulnerable to the exploitation of public goods by selfish individuals who contribute nothing to the joint venture \cite{axelrod1981evolution,nowak2004emergence,Hilbe18nature,chen_prsb_2019}. In this way, cooperation will eventually collapse \cite{hardin1968tragedy}. Conversely, cooperation described by providing public goods is universal ranging from microorganism to human society \cite{rainey2003evolution,sigmund2010calculus,perc2013inter,rand2013human,Broom2013,perc2017statistical}. Thus what drives the emergence of cooperative behaviour has attracted widespread attention from broad range of disciplines \cite{axelrod1981emergence,santos2005scale,Fu2008,Fu2009,huang2020int,Han2015}.

In past decades, a substantial body of theoretical and experimental investigations have illuminated various solutions including norms and incentive mechanisms that can be used to promote the evolution of cooperation \cite{boyd1992punishment,sigmund2001reward,milinski2002reputation,Hauert2002Volunteering,boyd2003evolution,andreoni2003carrot,fehr2004social,fehr2004third,Herrmann2008,Tavoni2011,santos2018social,Hilbe2018}. A frequently discussed incentive mechanism is punishment, with which some cooperators pay extra cost to punish selfish individuals \cite{fehr2000cooperation,fehr2002altruistic,hilbe_prsb_2010,zhang2013jtb,chen_14_jrsi,han_16_ab,han_21_plos}. As a special form of punishment, social exclusion has been introduced. Indeed this incentive mechanism has been found in realistic human and biological interactions~\cite{Melis2006Science,Wang2014PNAS,zhao2019BMC,Pentz2020}. For instance, cooperating groups of bacteria (wild-type Pseudomonas aeruginosa) growing on protein can restrict cheater emergence by producing costly cyanide to exclude LasR-null social cheaters from sharing the provided public goods~\cite{Wang2014PNAS,zhao2019BMC}. Chimpanzees can exclude non-cooperative partners in multiple interactions \cite{Melis2006Science}. In addition, in human society ranging from hunter-gatherer to industrialized societies, it is a common phenomenon that individuals are willing to choose cooperative partners and to isolate free-riders \cite{Martinez_17_sr,Bj2018Incidental,Martinez_15_sr}. Recently, social exclusion has been investigated from the theoretical perspective \cite{sasaki2013evolution}. It is often assumed that free-riders will be expelled from the beneficiaries and cannot get any benefit when exclusion is implemented, which is different from the setting of traditional punishment under which free-riders are imposed with a fixed penalty fine. Social exclusion strategy has been found to be more favourable to the evolution of cooperation compared with the traditional punishment, since its introduction into evolutionary games \cite{li2015social,liu2017competitions,szolnoki2017alliance}.

Although the aforementioned studies have enhanced our understanding about how social exclusion influences cooperative behaviour, they generally build on some limiting and questioning assumptions. To be specific, the evolutionary dynamics of exclusion strategy are always explored in ephemeral one-shot, but not repeated interactions. Furthermore, excluders only need to pay the exclusion cost for excluding defectors. Indeed in previous behavioural experiments about social exclusion, it is often required that excluders in the group must be able to identify ``bad apples" before excluding them in repeated public goods games \cite{masclet2003ostracism,cinyabuguma2005cooperation,kerr2009many}. These excluders identify cooperators and defectors by recording players' names and allocation decisions, which could require excluders to pay the extra cost of monitoring and information collecting in the interaction process. Hence when the above mentioned restrictions are released, it is unclear whether exclusion is still a viable strategy and how it can facilitate cooperation from the theoretical perspective, which are worth investigating.

In this work, we thereby propose a social exclusion strategy and explore its evolutionary consequence in repeated public goods games. We assume that excluders can monitor the decisions of all group members permanently to identify free-riding behaviours in the repeated interaction process by paying a monitoring cost, and subsequently expel these identified free-riders from group benefits. Our results reveal that the exclusion round is crucial for cooperation to thrive, since it determines how much defectors can benefit from previous repeated interactions. In infinite well-mixed populations, we observe evolutionary oscillations among cooperators, defectors, and excluders when free-riders can be expelled at the earliest stage. Similarly, stochastic evolution in finite populations can favour cooperation over defection when implemented exclusion is timely.

\vbox{}
\noindent\textbf{2. Model and Methods}

\noindent We consider a well-mixed population from which $N$ individuals are selected randomly to form an interaction group for playing a repeated public goods game. There are three game strategies: cooperation ($C$), defection ($D$), and peer exclusion ($E$). Both $C$ and $E$ individuals contribute $c$ to the common pool in every round, while $D$ individuals do nothing. In every round the accumulated contributions are multiplied by an enhancement factor $F$ ($1<F<N$) and subsequently shared equally among all group members, irrespective of whether they contributed or not. In the framework of repeated group interactions, the above game process will be repeated with probability $w$ ($0<w<1$), resulting in an average number of $\langle r\rangle=\frac{1}{1-w}$ rounds~\cite{van2012emergence}. Here we assume that peer excluders pay a permanent cost $\sigma$ to monitor the game process for identifying $D$ individuals and subsequently exclude all identified defectors from the group at the $\varsigma$-th round by paying an additional cost $c_{E}$. The excluded free-riders cannot share the public goods of the $\varsigma$-th round and will lose the opportunity to participate in the following possible game rounds.

It is worth noting that if the average number of rounds is smaller than the number of round in which defectors are excluded, namely, $\langle r\rangle < \varsigma$ , then peer exclusion strategists completely degenerate into pure cooperators except the permanent cost $\sigma$ of monitoring. Thus in this case defection is a more advantageous strategy than peer exclusion and cooperation~\cite{sasaki2013evolution}. Accordingly, we can write the payoff of a cooperator, defector, and excluder playing the repeated public goods game in a group in which $N_{C}$ cooperators, $N_{E}$ excluders, and $N_{D}$ defectors are present. Thus the related payoff values for cooperators, defectors, and excluders are respectively give as

\setlength{\arraycolsep}{0.0em}
\begin{eqnarray}\label{eq12}
\pi_{C}&=&\left\{
\begin{aligned}
&\frac{Fc(N_{C}+N_{E}+1)}{N}(\varsigma-1)+Fc(\langle r\rangle-\varsigma+1)-\langle r\rangle c,  \quad  \text{if} \ \varsigma \leq \langle r\rangle \quad\text{and}\quad N_{E}\neq0;\\
&\frac{Fc(N_{C}+N_{E}+1)}{N}\langle r\rangle-\langle r\rangle c,  \quad  \text{otherwise}.
\end{aligned}
\right.\\\label{eq22}
\pi_{D}&=&\left\{
\begin{aligned}
&\frac{Fc(N_{C}+N_{E})}{N}(\varsigma-1),  \quad  \text{if} \ \varsigma \leq \langle r\rangle \quad\text{and}\quad N_{E}\neq0;\\
&\frac{Fc(N_{C}+N_{E})}{N}\langle r\rangle,  \quad  \text{otherwise}.
\end{aligned}
\right.\\\label{eq32}
\pi_{E}&=&\left\{
\begin{aligned}
&\frac{Fc(N_{C}+N_{E}+1)}{N}(\varsigma-1)+Fc(\langle r\rangle-\varsigma+1)-\langle r\rangle c-c_{E}N_{D}-\sigma,     \quad   \text{if} \ \varsigma \leq \langle r\rangle;\\
&\frac{Fc(N_{C}+N_{E}+1)}{N}\langle r\rangle-\langle r\rangle c-\sigma,  \quad  \text{otherwise}.
\end{aligned}
\right.
\end{eqnarray}
\setlength{\arraycolsep}{5.0pt}

We then apply evolutionary game theory to investigate the evolution of strategies in our model~\cite{sigmund2010calculus} . To be specific, we will respectively consider our model in infinite and finite well-mixed populations in the following.\\

\noindent \textbf{(a) Infinite well-mixed populations.}\\
We consider an infinite well-mixed population, in which $N$ individuals are selected randomly to engage in the repeated public goods game. Since random sampling leads to groups with compositions that follow a multivariate binomial distribution, the average payoff $P_{i}$ of one individual adopting a given strategy $i$ can be calculated as $$P_{i}=\sum_{N_{C}=0}^{N-1}\sum_{N_{D}=0}^{N-N_{C}-1}\binom{N-1}{N_{C}}\binom{N-N_{C}-1}{N_{D}}x^{N_{C}}y^{N_{D}}z^{N-N_{C}-N_{D}-1}\pi_{i},$$
where $i=C$, $D$, or $E$.

In order to investigate the evolutionary dynamics in infinite well-mixed populations, we use the replicator equation approach~\cite{hofbauer1998evolutionary,hofbauer2003evolutionary}. To do that, we use $x, y,$ and $z$ to denote the fraction of $C, D$, and $E$ players, respectively. Correspondingly, we have $x + y + z = 1$ and the replicator equation can be written as
\begin{eqnarray}\label{rep}
\left\{
\begin{aligned}
\dot{x}&=x(P_{C}-\bar{P}),\\
\dot{y}&=y(P_{D}-\bar{P}), \\
\dot{z}&=z(P_{E}-\bar{P}),
\end{aligned}
\right.
\end{eqnarray}
where $\bar{P}=xP_{C}+yP_{D}+zP_{E}$ represents the average payoff of the whole population. More detailed theoretical analysis of the replicator dynamics can be found in Electronic Supplementary Material.

Furthermore, since replication is error-prone~\cite{nowak2001evolution,page2002unifying}, we also incorporate mutation into the above deterministic replicator dynamics. The resulting replicator-mutator equation can be written as
\begin{eqnarray}
\left\{
\begin{aligned}
\dot{x}&=xP_{C}q_{C\rightarrow C}+yP_{D}q_{D\rightarrow C}+zP_{E}q_{E\rightarrow C}-x\bar{P}, \\
\dot{y}&=yP_{D}q_{D\rightarrow D}+xP_{C}q_{C\rightarrow D}+zP_{E}q_{E\rightarrow D}-y\bar{P},\\
\dot{z}&=zP_{E}q_{E\rightarrow E}+xP_{C}q_{C\rightarrow E}+yP_{D}q_{D\rightarrow E}-z\bar{P},
\end{aligned}
\right.
\end{eqnarray}
where $q_{U\rightarrow V}$ denotes the probability that strategy $U$ generates an offspring using strategy $V$ ($U, V=C, D$, or $E$). A detailed description and analysis of the replicator-mutator equation is present in Electronic Supplementary Material.\\

\noindent \textbf{(b) Finite well-mixed populations.} \\
\noindent We further consider our model in a finite well-mixed population in which the population size is $Z$. For finite well-mixed populations of size $Z$ with $i_{C}$ cooperators, $i_{E}$ excluders, and $Z-i_{C}-i_{E}$ defectors, we know that the average payoffs of these three strategies in a configuration $\textbf{i}=\{i_{E},i_{C}\}$ can be respectively computed by using a hypergeometric sampling~\cite{Hauert2002Volunteering}. More precisely, the average payoffs of cooperators, defectors, and excluders in the configuration $\textbf{i}=\{i_{E},i_{C}\}$ can be respectively given as
\begin{eqnarray}\label{avepay}
f_{C}&=&\sum_{N_{C}=0}^{N-1}\sum_{N_{E}=0}^{N-N_{C}-1}\frac{\binom{i_{C}-1}{N_{C}}\binom{i_{E}}{N_{E}}\binom{Z-i_{C}-i_{E}}{N-N_{C}-N_{E}-1}}{\binom{Z-1}{N-1}}\pi_{C},\nonumber\\
f_{D}&=&\sum_{N_{C}=0}^{N-1}\sum_{N_{E}=0}^{N-N_{C}-1}\frac{\binom{i_{C}}{N_{C}}\binom{i_{E}}{N_{E}}\binom{Z-i_{C}-i_{E}-1}{N-N_{C}-N_{E}-1}}{\binom{Z-1}{N-1}}\pi_{D},\\
f_{E}&=&\sum_{N_{C}=0}^{N-1}\sum_{N_{E}=0}^{N-N_{C}-1}\frac{\binom{i_{C}}{N_{C}}\binom{i_{E}-1}{N_{E}}\binom{Z-i_{C}-i_{E}}{N-N_{C}-N_{E}-1}}{\binom{Z-1}{N-1}}\pi_{E},\nonumber
\end{eqnarray}
where $\pi_{C}, \pi_{D}$, and $\pi_{E}$ respectively denote the payoffs of $C, D$, and $E$ individuals obtained from the game, which are shown by Eqs.~(\ref{eq12})-(\ref{eq32}).

Subsequently, we consider the pairwise comparison rule combined with mutation to describe how the number of individuals adopting a given strategy evolves in finite well-mixed populations \cite{imhof2005evolutionary,vasconcelos2013bottom,vasconcelos2014}. To be specific, with probability $\mu$, mutation occurs and one randomly selected player $L$ picks a new strategy randomly from the remaining available strategies space. With probability $1-\mu$, player $L$ imitates the strategy of another randomly selected player $R$ according to $\bar{p}=\frac{1}{1+\exp{[\beta(f_{L}-f_{R})]}},$
where $\beta\geq0$, the so-called intensity of selection, translates into noise associated with errors in the imitation process \cite{szabo_pre98}. Based on the above description, we can write the transition probability that an individual selected from $i_{U}$ players with a given strategy $U$ adopts another different strategy $V$ as \cite{vasconcelos2013bottom}
\begin{eqnarray*}
T_{U\rightarrow V}=(1-\mu)\big[\frac{i_{U}}{Z}\frac{i_{V}}{Z-1}\frac{1}{1+\exp{(\beta(f_{U}-f_{V}))}}\big]+\mu\frac{i_{U}}{(d-1)Z},
\end{eqnarray*}
where $U,V=C, D,$ or $E$, and $d$ is the number of alternative strategies in the strategy space.

For arbitrarily mutation rates, the evolutionary dynamics among $C, D,$ and $E$ can be described by an embedded Markov process over a two-dimensional space. Transitions take place between different configurations of the system characterized by the vector $\textbf{i}(t)=\{i_{E}, i_{C}\}$. The study of Markov process is to mainly determine the evolution of its probability density function, $p_{\textbf{i}}(t)$, which provides information on the prevalence of each configuration at time $t$ \cite{vasconcelos2013bottom,vasconcelos2014}. The transition probability and probability density function obey the discrete time Master Equation \cite{kampen2007}. The so-called stationary distribution $\bar{p_{\textbf{i}}}$ is obtained by reducing the master equation to an eigenvector search problem. More details about methods and theoretical analysis can be found in Electronic Supplementary Material.

By means of individual-based simulations, we also investigate social learning dynamics with an arbitrary exploration rate $\mu$ in finite populations \cite{Sigmund2010Social}. Each individual obtains an average payoff calculated by Eq.~(\ref{avepay}) based on the random sampling of the interaction groups. Strategies evolve via a mutation-selection process that can be used to describe the evolution of strategies in discrete time. At each time step, one player $L$ is randomly selected to update its strategy in the following way: with probability $\mu$, player $L$ undergoes a mutation and randomly adopts one of the available strategies. With probability $1-\mu$, another player $R$ is selected randomly to be a role model for player $L$. Then player $L$ imitates the strategy of player $R$ with the probability described by Fermi function $\bar{p}=\frac{1}{1+\exp{[\beta(f_{L}-f_{R})]}}$.

\vbox{}
\noindent\textbf{3. Results}

\noindent In the following, we respectively present our results of evolutionary dynamics of cooperators, defectors, and excluders in infinite and finite well-mixed populations.\\

\begin{figure*}
\centering
\includegraphics[width=.9\linewidth]{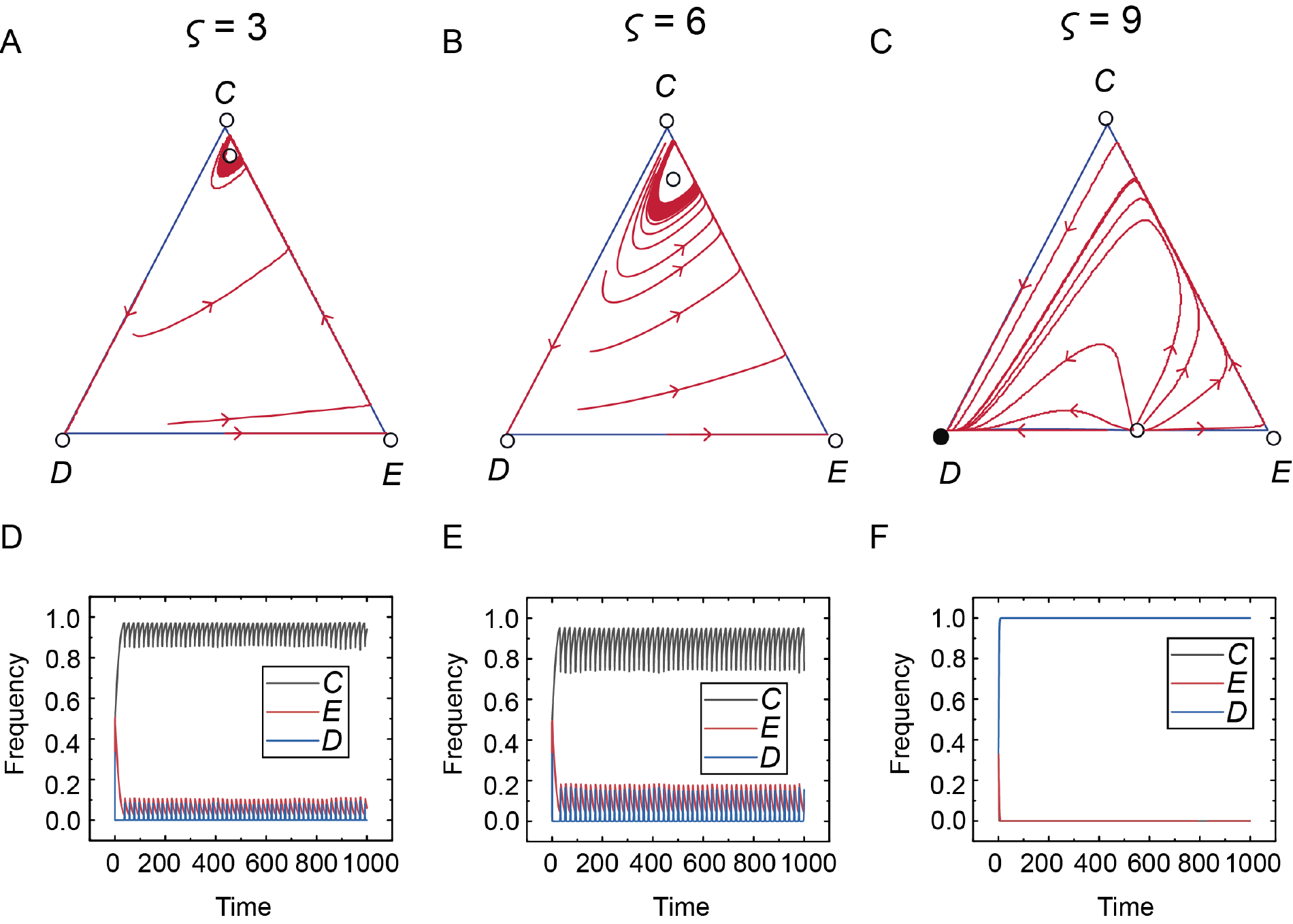}
\caption{Deterministic replicator dynamics among cooperators, defectors, and excluders in an infinite well-mixed population with different exclusion rounds $\varsigma$.
The triangle represents the state space $S_{3}=\{(x, y, z): x, y, z\geq0, x+y+z=1\}$, where $x, y,$ and $z$ denote the frequencies of cooperators, defectors, and excluders, respectively. The three vertices $C, D$,
and $E$ correspond to the three homogeneous states in which the population exclusively consists of cooperators $(x = 1)$, defectors $(y = 1)$, or excluders $(z = 1)$. The open circles represent unstable equilibria, and the filled circles denote stable equilibria. Panels D-F show the time evolution of fractions of three strategies $C$ (black line), $E$ (red line), and $D$ (blue line) for three different $\varsigma$ values. Parameter values are $N=5, F=3, c=1, c_{E}=0.4, w=0.9,$ and $\sigma=0.1$.}
\label{fig1}
\end{figure*}

\noindent \textbf{(a) Evolutionary dynamics in infinite well-mixed populations.}\\
\noindent Let us first focus on the evolutionary dynamics of the three strategies $C$, $D$, and $E$ described by Eq.~(\ref{rep}) in infinite well-mixed populations. According to the theoretical analysis, we distinguish three different cases where the outcome changes with increasing exclusion round $\varsigma$. For small exclusion round satisfying $\varsigma < \frac{N[(F-1)\langle r\rangle c-\sigma-(N-1)c_{E}]}{(N-1)Fc}+1$, cooperation, defection, and exclusion can form a cyclic dominance that resembles a rock-scissors-paper cycle on the boundary of $S_{3}$, as shown in Fig.~\ref{fig1}A and B. Here the evolution on the $CD$ edge is unidirectional from $C$ to $D$. It goes from $D$ to $E$ on the $DE$ edge and goes from $E$ to $C$ on the $EC$ edge. In Electronic Supplementary Material, we theoretically prove that there is an asymptotically stable heteroclinic cycle on the boundary of $S_{3}$. We show that there exists a limit cycle in the interior of the simplex, in which all interior trajectories of the state space form evolutionary oscillations around the interior equilibrium when $\varsigma$ is small (see Fig.~\ref{fig1}A, B, D, and E). As $\varsigma$ increases and satisfies $\frac{N[(F-1)\langle r\rangle c-\sigma-(N-1)c_{E}]}{(N-1)Fc}+1 < \varsigma < \frac{N[(F-1)\langle r\rangle c-\sigma]}{(N-1)Fc}+1$, an unstable equilibrium appears on the $DE$ edge and the state where each individual chooses to defect is the only stable equilibrium. Numerical calculations confirm that all trajectories started in the state space converge to a state where all $D$ players appear (see Fig.~\ref{fig1}C and F). As $\varsigma$ further increases and exceeds $\langle r\rangle$, defectors have more advantages over both cooperators and excluders, so that the sate of full $D$ is globally stable.

Next, we study the effects of other parameters on the replicator dynamics of cooperators, defectors, and excluders, as shown in Electronic Supplementary Material. We find that cooperative behaviours can be maintained in the population when the exclusion round $\varsigma$ is not too large. Particularly, in the specific case of $\varsigma=1$ we can reproduce previous findings that excluders can emerge in a sea of defectors and dominate them \cite{sasaki2013evolution}. Furthermore, we provide some numerical examples to show the effects of different observation cost $\sigma$ (Fig.~S2) and discount factor $w$ (Fig.~S4) on replicator dynamics. We find that our main results remain valid if the value of $\sigma$ is approximately changed. Besides, the cyclic dominance among cooperators, defectors, and excluders can still appear when the value of $w$ is not too small.

\begin{figure*}
\centering
\includegraphics[width=.9\linewidth]{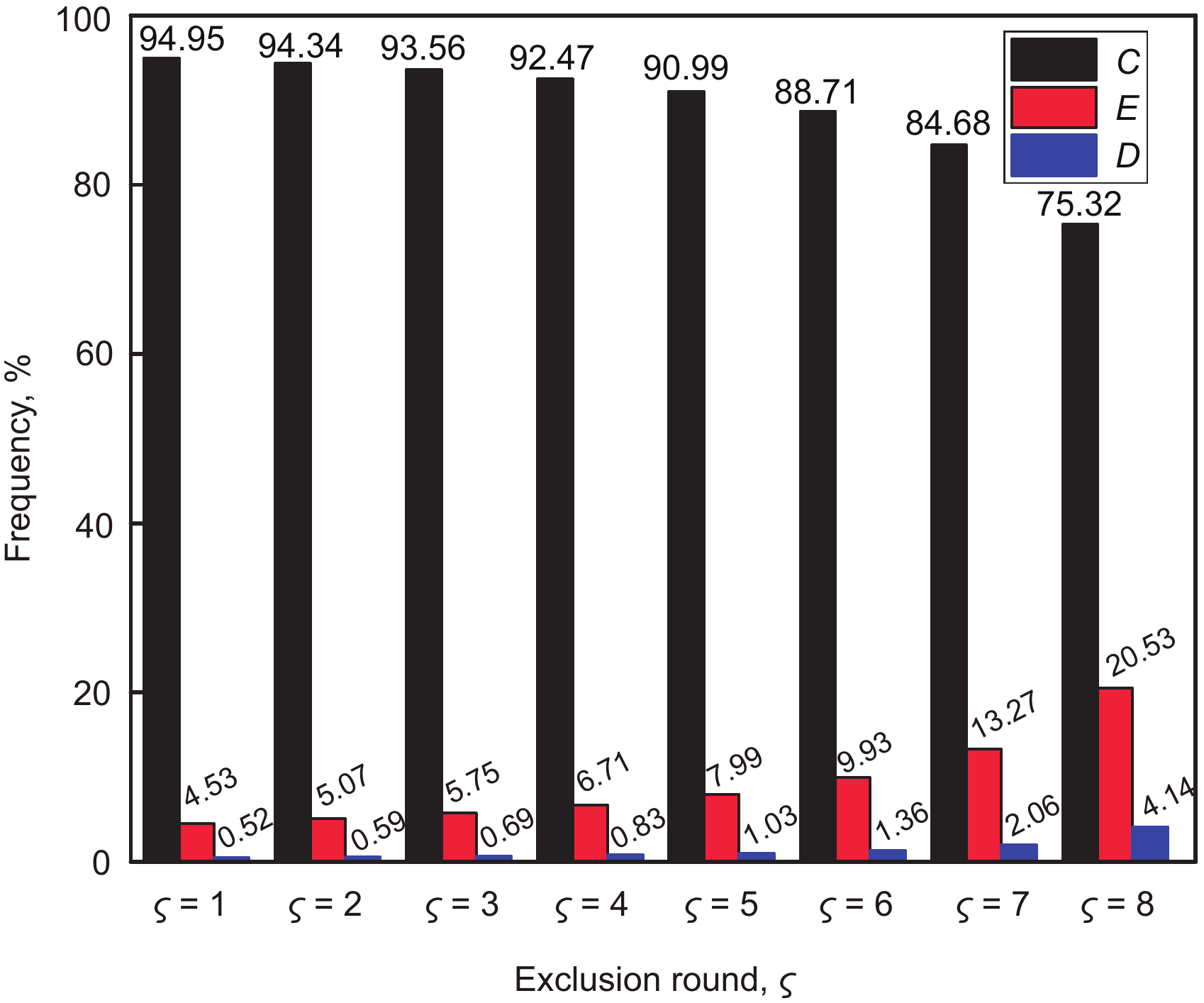}
\caption{Average levels of cooperation, defection, and exclusion for different $\varsigma$ values. When $C, D,$ and $E$ can form an evolutionary oscillation, the average level of each strategy can be calculated over an approximate periodic orbit. These frequencies of $C, D$, and $E$ strategies vary with different exclusion round values $\varsigma$, as indicated on the horizontal axis. Parameter values are $N = 5, F = 3, c = 1, c_{E}=0.4, w=0.9,$ and $\sigma=0.1$.}
\label{fig2}
\end{figure*}

In order to shed light on the details of the evolutionary oscillations dynamics shown in the Fig.~\ref{fig1}A and B, we depict the average levels of the mentioned three strategies over an approximate periodic orbit for different exclusion rounds. Fig.~\ref{fig2} illustrates that the highest cooperation rates can be obtained for immediate exclusion. The intuitive explanation is that the maximal cooperation level arises when excluders can quickly expel free-riders from the beneficiaries. Interestingly, although the average level of pure cooperators decreases with the increase of $\varsigma$, it is always higher than that of excluders and free-riders, as long as the exclusion is executed not too late.

Furthermore, we extend the replicator equation and consider the ``replicator-mutator equation" for frequency-dependent selection with mutation in infinite well-mixed populations. For small mutation rate, we find that there is also an unstable interior equilibrium containing all three strategies, and an evolutionary oscillation forms around this equilibrium (Fig.~S5A in Electronic Supplementary Material). If the mutation rate exceeds a critical value, the evolutionary oscillation disappears and all interior trajectories converge to a stable fixed point. With the increase of $\mu$, the interior equilibrium point moves towards the inner part of simplex gradually (Fig.~S5B-E in Electronic Supplementary Material).\\

\noindent \textbf{(b) Evolutionary dynamics in finite well-mixed populations.}\\
\noindent We note that replicator equations cannot be used directly to describe the evolutionary dynamics in a more realistic system where the population size is finite. In the latter case, stochastic effects including behavioural mutations or errors of imitation may play important roles in the evolutionary dynamics \cite{imhof2005evolutionary,Vasconcelos2017Stochastic}. In the following, we investigate the stochastic dynamics of finite populations when mutation rates are arbitrarily large and sufficiently small.

\begin{figure*}
\centering
\includegraphics[width=.9\linewidth]{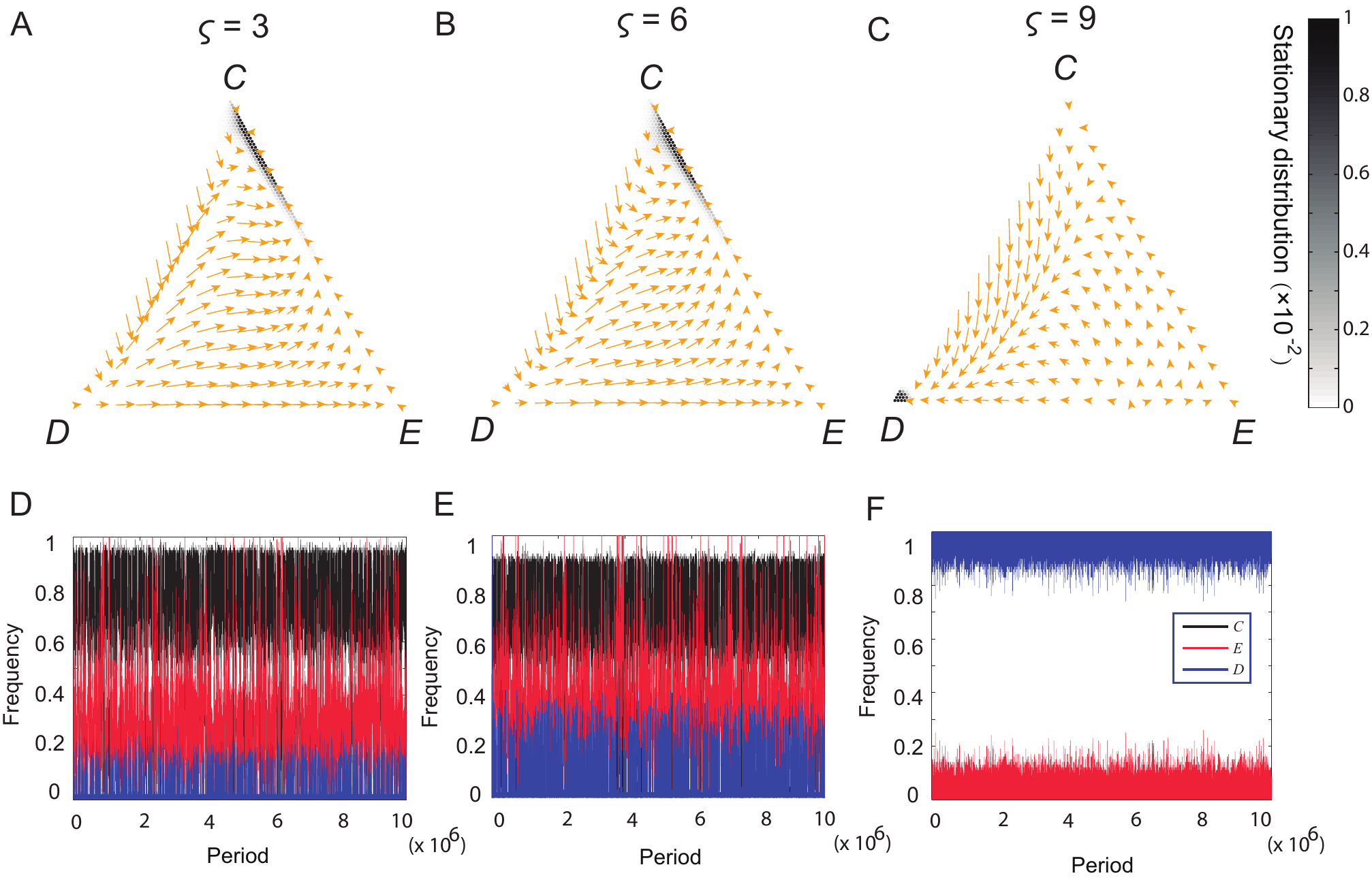}
\caption{Evolutionary dynamics of cooperation, defection, and exclusion in finite well-mixed populations for different exclusion round $\varsigma$ values. The dynamics are best characterized by the full stationary distribution, which is mapped onto the triangular simplex. Each simplex contains all possible configurations of the population, and each state is denoted by a small cycle. The value of the stationary distribution at each configuration is shown using a grey scale, where darker dots denote those configurations in which the population spends more time. The red arrows depict the so-called gradient of selection, which provide the most probable directions of evolution. Panels~D-F show the time evolution of strategies by individual-based simulations for different $\varsigma$ values. Parameter values are $Z=100, N=5$, $F=3$, $c=1$, $c_{E}=0.4, \beta=2, \sigma=0.1, \mu=0.01,$ and $w=0.9$.}\label{fig3}
\end{figure*}

When mutation rate is arbitrarily large, the stochastic effects can drive the system from the vicinity of one configuration of the state space to another. The stochastic dynamics associated with these three strategies are best characterized by the gradient of selection and the full stationary distributions, as shown in Fig.~\ref{fig3}A-C and Figs.~S8-S10. Each triangular simplex contains all possible configurations of the finite population where each configuration is represented by a small circle. The magnitude of stationary distributions is shown by using a grey scale where darker areas indicate more often visited configurations. The arrows show the so-called gradient of selection, which provides the most likely direction of evolution from a given configuration, which can be calculated from the drift term of the Fokker-Planck equation \cite{Helbing1993Boltzmann}.

Similarly to the evolutionary outcomes in infinite well-mixed populations, the qualitative behavioral dynamics in the finite population change with the exclusion round $\varsigma$. As shown in Fig.~\ref{fig3}A-C, three representative examples of the behavioural dynamics among $C, D$, and $E$ players for different exclusion rounds $\varsigma$ are presented when mutation is not negligible. For small exclusion rounds, if most players of the population are $C$ players, it is better to be a $D$ player due to social dilemma. If $D$ players are prevalent, $E$ players can rapidly outcompete $D$ players, leading the population to the configuration that contains all-$E$ players. In the latter case, behavioural mutations allow the emergence of $C$ players in the population and $E$ players have less advantage than $C$ players since $C$ players do not need to bear additional monitoring costs. Accordingly, $C$ players spread. As a result, a cyclical evolutionary outcome can be formed (see the direction of arrow flow). This outcome is also verified by individual-based simulations (see Fig.~\ref{fig3}D and E). Besides, as shown by the stationary distributions, the population will spend most of the time in configurations comprising a sizable amount of $C$ players together with $E$ players, which can prevent $D$ players from invading (see Fig.~\ref{fig3}A and B).

For high values of exclusion round ($\varsigma=9$), the behavioural dynamics are quite different. As shown in Fig.~\ref{fig3}C, the population spends a significant amount of time in configurations in which defectors spread. Fig.~\ref{fig3}E also reveals that the whole population is taken over completely by defectors. It is not surprising that such outcome occurs when the value of exclusion rounds is too large because $D$ players have obtained a higher payoff before they are excluded from the repeated group interactions by $E$ players. Furthermore, exclusion strategy will be completely ineffective in resisting defection when exclusion round exceeds the average game rounds, allowing the population to spend longer periods of time in configurations with many defectors.

Finally, we present theoretical and numerical results for the stationary distribution when the mutation rate is sufficiently small. In this case, the time scales of mutation and imitation are separated, thus the population state is homogeneous most of the time. The evolutionary dynamics are determined by an embedded Markov chain where the state space is composed of homogeneous states (more details can be found in Electronic Supplementary Material). Both theoretical and numerical results underline that exclusion strategy can have more evolutionary advantages over the other two strategies for weak selection when the exclusion round is small (Fig.~S10A). For strong selection, however, the long-run frequencies of the three competing strategies are identical (Fig.~S11A). A detailed analytical approximations of the transition matrix among these three homogeneous states shows that the long-run frequencies in the ($C, D, E$) subpopulations are $(\frac{1}{3},\frac{1}{3},\frac{1}{3})$ in the strong selection limit (see Electronic Supplementary Material).

Similarly to the above described cases, when exclusion round is high, defection can have more advantages than the two other strategies for weak selection (Fig.~S10B). With the increase of intensity of selection, the advantage of defectors increases gradually. In the strong selection, defection is the most advantaged strategy, yielding $(0, 0, 1)$ fractions for $C, E$, and $D$ strategies, respectively (Fig.~S11B). A detailed theoretical analysis of the transition matrix of the system and the linear approximation of the stationary distribution in the weak selection limit can be found in Electronic Supplementary Material.

\begin{figure*}
\centering
\includegraphics[width=.9\linewidth]{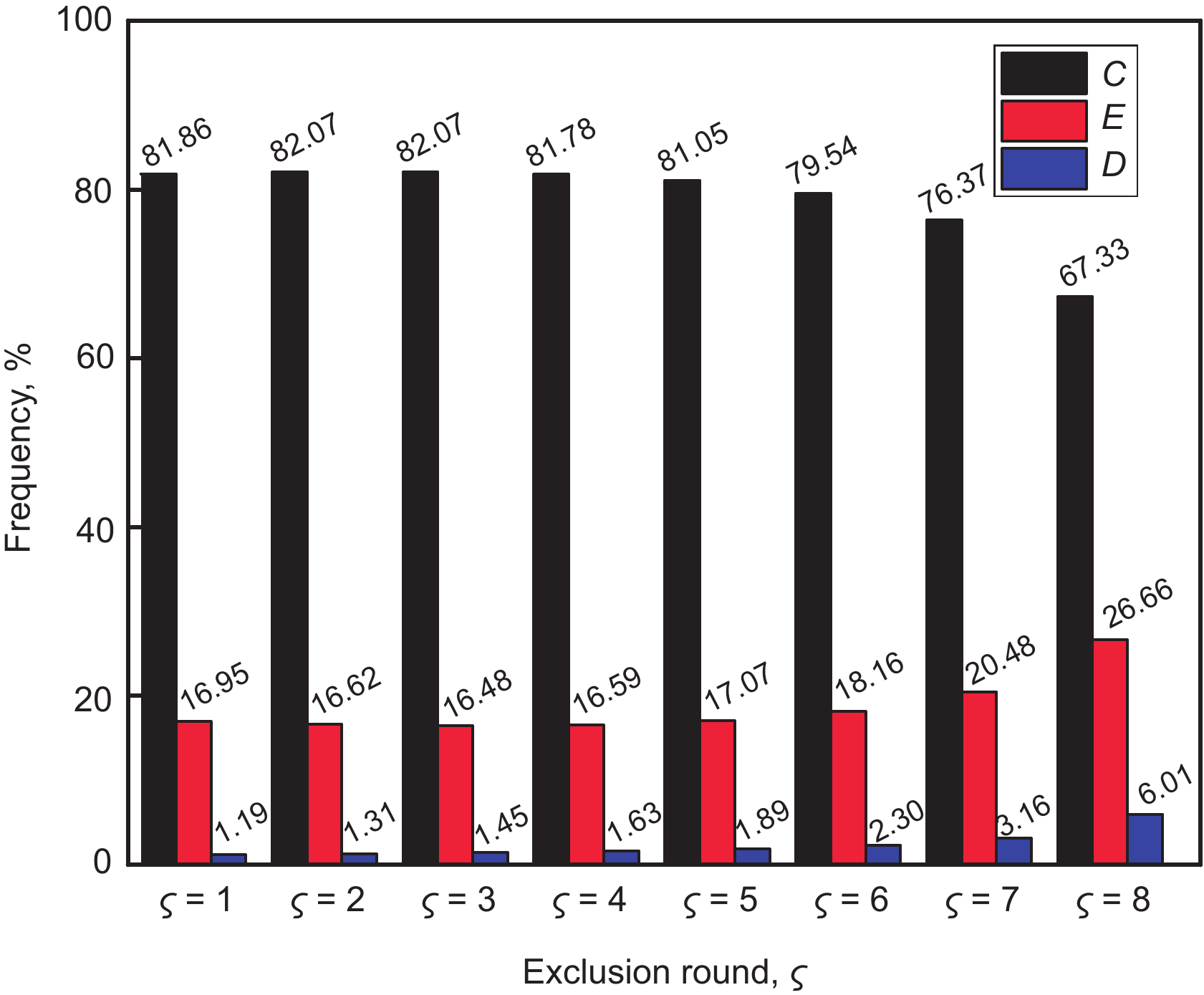}
\caption{The average levels of cooperation, defection, and exclusion in finite populations for different exclusion round values $\varsigma$. These frequencies of $C, D$, and $E$ strategies vary with different exclusion round values $\varsigma$, as indicated on the horizontal axis. Parameter values are $Z=100, N=5$, $F=3$, $c=1$, $c_{E}=0.4, \beta=2, \sigma=0.1, \mu=0.01$, and $w=0.9$.}\label{fig4}
\end{figure*}

In Fig.\ref{fig4}, we further show the average levels of cooperation, exclusion, and defection for different values of $\varsigma$ when mutation cannot be ignored. Interestingly, the average cooperation level slightly increases first then decreases with the increase of the exclusion round $\varsigma$. The changing trend about the cooperation level is opposite to that about the average level of excluders. Besides, the level of defection increases with the increase of $\varsigma$. Altogether, Fig.~\ref{fig4} confirms that cooperation can be maintained at a high level as long as free-riders can be excluded in the early rounds. Indeed, in the presence of mutation, the population will never fixate in any of the absorbing states. In addition, the intensity of selection incorporates noise associated with errors into the process of strategy updates. Accordingly, we can find that there exists evolutionary oscillations for the frequencies of cooperators, defectors, excluders, as shown in Fig.~\ref{fig3}D and E. Since the cooperation level shown in  Fig.\ref{fig4} averaging over all possible configurations are weighted with the corresponding strategy distribution, the oscillation effects has been involved and we can see that the cooperation level slightly increases with the number of rounds even when the values of intensity of selection and mutation rate are changed appropriately. However, we stress that our main conclusion that immediate exclusion is favourable to the evolution of cooperation is still valid in finite well-mixed populations.

\vbox{}
\noindent\textbf{4. Discussion}

\noindent In modern societies, it has been widely observed that ostracism or the exclusion of disapproved members from groups, teams, and communities can enforce norm conformity and cooperation \cite{Bj2018Incidental,zhao2019BMC}. However, it still needs further theoretical efforts to understand its evolutionary advantage. In this work, we have developed an evolutionary game model to explore the dynamical consequences of social exclusion strategy in the public goods game where group members are engaged in repeated interactions. We assume that excluders need to monitor the whole group interactions for collecting relevant information before excluding free-riders. If the exclusion is executed, then free-riders cannot obtain the benefits of common pool anymore, but they can have the incomes originated from previous rounds. We have found that the actual round at which defectors are excluded plays a decisive and highly non-trivial role in the general well-being. Interestingly, the evolutionary oscillation among the competing strategies can be detected if the exclusion is executed at an early stage in repeated interactions. More surprisingly, the time averages of the temporary fractions of strategies display the outcome of the dominance of cooperators. We have also investigated the stochastic dynamics of a finite population in the presence of behavioural mutations and found that immediate exclusion can lead to a high level of cooperation.

In this work, we investigate the effects of ostracism on cooperation in the multi-period public goods game by comparing the treatments with and without exclusion, mainly inspired by some interesting behavioural experiments \cite{masclet2003ostracism,cinyabuguma2005cooperation,kerr2009many,maier2010ostracism}. In the former experiments, subjects are given the opportunity to exclude each other after they are informed about the individual contributions of other group members, and player who is ostracized cannot take part in the subsequent interactions. The experimental results show that subjects exhibit a willingness to exclude other members of their group and the threat of social exclusion for uncooperative behaviour can significantly improve the level of contribution \cite{masclet2003ostracism}. It is worth emphasizing that two hypotheses have been considered to identify uncooperative ``bad apples" in previous experiment: cooperation decisions are not anonymous; group members can be identified by name \cite{kerr2009many}. Inspired by the above-mentioned behavioural experiments, our mathematical model considers that excluders need to pay a permanent cost of monitoring to identify other players' behaviours during the repeated group interaction process. Indeed, such monitoring cost also called opportunity cost in the previous work \cite{han_csr21}, has been used to describe the transparency between individuals in game interactions \cite{lehrer_te18}. This realistic set-up makes our model more reasonable and brings a step closer towards understanding how social exclusion strategy works in the realistic case.  Accordingly, our work is essentially different from the previously reported incidental ostracism \cite{Bj2018Incidental}. To be specific, the ways of implementing social exclusion and incidental ostracism are different. Previous study assumes that individuals can select partners and those who are not selected are excluded from social interactions \cite{Martinez_17_sr,Bj2018Incidental,Martinez_15_sr}. However, in our work we regard social exclusion as instrumental punishment of free-riders. In addition, participants play an iterated prisoner's dilemma or coordination game with partner choice in the previous study~\cite{Bj2018Incidental}. While in our model, we explore the evolutionary dynamics of social exclusion strategy in repeated group interactions. Our results underline that cooperative behaviour cannot be promoted when implementing exclusion on free-riders is too late, which is consistent with the experimental results that the introduction of ostracism increases contribution levels except in the last period \cite{maier2010ostracism}. Our calculations also reveal that too early exclusion could be a source of modest cooperation.

The positive role of peer exclusion in the evolution of cooperation has already been reported in previous theoretical model developed in large well-mixed populations \cite{sasaki2013evolution}. However, the mentioned work mainly examined the dynamics of the system when interaction among group members is one-shot and defectors can be detected in a cost-free manner. In our model, we release these limitations. In addition, in our model we assume that social exclusion are implemented simultaneously and faultlessly in all groups and we do not consider the case in which defectors will be excluded from the group probabilistically. The main reason is that in our model we consider social excluders have paid the observation cost in advance to identify all the free-riders during the repeated group interaction process. Interestingly, we explore new dynamical behaviours, such as evolutionary oscillations, as summarized in Fig.~\ref{fig1}. It is worth noting that evolutionary oscillations can be also observed in the interactions of Pseudomonas aeruginosa when Pseudomonas aeruginosa uses the strategy of social exclusion by using cooperator-released toxic cyanide to exclude lasR mutant defectors lacking the LasR regulon for detoxification~\cite{Wang2014PNAS,zhao2019BMC}. Besides, we can reproduce previous findings that rare excluders can emerge in a sea of defectors and subvert them~\cite{sasaki2013evolution} (also see Fig.~S1). Our work may unveil the effects of social exclusion on the evolution of cooperation in a realistic manner.

As we have found, the application of different exclusion rounds can induce rather distinct dynamical behaviours. We stress that the actual number of exclusion rounds is not the only factor that affects evolutionary dynamics. As Fig.~S2 in Electronic Supplementary Material demonstrates clearly, the cost of monitoring can modify the amplitude of oscillation significantly, thus altering the evolutionary advantage of this incentive strategy. Another important parameter is the discount factor of repeated games, which strongly influences the cooperation level. Our results highlight that cooperative behaviours cannot emerge when the mentioned discount factor is lower than an intermediate value (Fig.~S4). Here, we show that it is beneficial for excluders to manage to expel free-riders from the beginning even in situations with high monitoring cost or intermediate discount factor. What's more, the noise or error that individuals can make during the evolutionary process might strongly influence outcomes of repeated games \cite{sigmund2010calculus}. As shown in Fig.~S5, replicator-mutator dynamics in infinite populations reveal that cooperators, defectors, and excluders can form evolutionary oscillations when the mutation rate is small. With the increase of mutation rate, the evolutionary oscillations disappear, and the three strategies can coexist stably in the population. Furthermore, stochastic mutation-selection dynamics demonstrate that the population spends most of the time in states which are close to the three corners of the simplex for early exclusion when mutation rate is small (see Fig.~S9A). While the population will spend increasing amounts of
time in states which are near the interior of the simplex for higher values of mutation rate (see Fig.~S9B). Our model thus predicts that quickly excluding free-riders can maintain a high level of cooperation over a wide range of parameters. Thus our observations may offer a potential way for how to use exclusion strategy efficiently in real and complicated situations.

In this work, we stress that once defectors are excluded from the interaction group, they cannot participate in the interactions any more in our model. In reality, it is a frequently applied practice that individuals who violate social rules can regain their qualification to participate in activities after a certain period of time. For example, a drunk driver whose driver's licence has been revoked can take a driving licence test again after some time. Also, this scenario has been applied to a behavioural experiment in which one player who is ostracized can re-enter the next round unless ostracized again \cite{Hirshlifer1989Cooperation}. For future study, thus a potential extension of our present work could be to explore the possible consequence in the realistic scenario where the finite-time effect is considered for exclusion of free-riders in repeated interactions. Moreover, it is meaningful and important to consider conditional strategies when repeated interactions are involved~\cite{Hilbe18nature,sigmund2010calculus,Hilbe2018}. Indeed previous studies have revealed that conditional punishment \cite{huang2018sr,Szolnoki2013jtb} or conditional exclusion \cite{quan2019chaos} can play important roles even in the one-shot public goods game. It is thus worth exploring the effect of conditional strategies on evolutionary dynamics of cooperation in repeated group interactions. Furthermore, individuals using conditional strategies may make different sorts of errors including strategy-implementation errors during the course of repeated interactions~\cite{chen_14_jrsi}, and thus it is worth further investigating the effects of errors when conditional strategies are considered in repeated group interactions for future study.

\noindent

\noindent \\ \textbf{Author contributions} \\
X.C., L.L., and A. S. designed the research, L.L. and Z.X. performed the research, X.C., L.L., and A. S.
wrote the manuscript, and all authors discussed the results and commented on and improved the
manuscript.

\noindent \textbf{Acknowledgments} \\
This work was supported by the National Natural Science
Foundation of China (Grant Nos. 61976048 and 62036002) and the Fundamental
Research Funds of the Central Universities of China.

\noindent \\ \textbf{Data availability}\\
The raw data generated with these computer calculations is available from the corresponding author
upon reasonable request.

\noindent \\ \textbf{Competing financial interests} \\
The authors declare no competing financial interests.

\newpage



\renewcommand{\figurename}{\textbf{Supplementary Figure}}

\centerline{Electronic Supplementary Material for}
\vbox{}

\centerline{\normalsize\textbf{Early exclusion leads to cyclical cooperation in repeated group interactions}}
\vbox{}

\centerline{Linjie Liu, Zhilong Xiao, Xiaojie Chen, and Attila Szolnoki}

\vbox{}

\noindent

\maketitle

\renewcommand{\figurename}{\textbf{Supplementary Figure}}



In Electronic Supplementary Material, we provide details of the analysis of evolutionary dynamics resulting from social exclusion strategy in repeated group interactions. Specifically, in Section~1 we present analytical results for our model in infinite well-mixed populations. In Section~2, we present detailed results of evolutionary dynamics for our model in finite well-mixed populations.

\vbox{}
\setcounter{section}{0}
\renewcommand\thesection{\arabic{section}}
\setcounter{subsection}{0}

\noindent\section{\textbf{Evolutionary dynamics in infinite well-mixed populations}}
\renewcommand\thesubsection{1.\arabic{subsection}}
\subsection{\textbf{Deterministic replicator dynamics}}

%

We assume an infinite well-mixed population, where $x$ denotes the fraction of cooperators ($C$), $y$ the fraction of defectors ($D$), and $z$ the fraction of peer excluders ($E$). A group of $N$ individuals are selected randomly from the whole population to participant in the public goods game (PGG) where all players engage in repeated group interactions. In the game, both $C$ and $E$ players contribute a certain amount $c$ to the common pool in every round, whereas $D$ players always opt for contributing nothing. The total contributions are then multiplied by a factor $F$ ($1<F<N$) and equally shared among all group members who participate in the PGG. This interaction process repeats itself with a probability $w$, resulting in an average number of $\langle r\rangle=\frac{1}{1-w}$ rounds for the interaction group. In the repeated interactions, $E$ players also need to pay a permanent cost $\sigma$ to monitor the whole game process and subsequently exclude all identified defectors from the group at the $\varsigma$-th round. When excluding defectors, they pay a cost $c_{E}$ on each defector in the group. Once defectors are excluded from the group, then they cannot enjoy the benefit of public goods anymore and will lose the opportunity to participate in the possible following rounds. As a result, the average payoffs of $C, D$, and $E$ players are respectively given as
\begin{eqnarray}\label{eq1}
P_{C}&=&\sum_{N_{C}=0}^{N-1}\sum_{N_{D}=0}^{N-N_{C}-1}\binom{N-1}{N_{C}}\binom{N-N_{C}-1}{N_{D}}x^{N_{C}}y^{N_{D}}z^{N-N_{C}-N_{D}-1}\pi_{C},\nonumber\\
P_{D}&=&\sum_{N_{C}=0}^{N-1}\sum_{N_{D}=0}^{N-N_{C}-1}\binom{N-1}{N_{C}}\binom{N-N_{C}-1}{N_{D}}x^{N_{C}}y^{N_{D}}z^{N-N_{C}-N_{D}-1}\pi_{D},\\
P_{E}&=&\sum_{N_{C}=0}^{N-1}\sum_{N_{D}=0}^{N-N_{C}-1}\binom{N-1}{N_{C}}\binom{N-N_{C}-1}{N_{D}}x^{N_{C}}y^{N_{D}}z^{N-N_{C}-N_{D}-1}\pi_{E},\nonumber
\end{eqnarray}
where $\pi_{C}, \pi_{D},$ and $\pi_{E}$ are, respectively, the payoffs of a $C, D$, and $E$ player in a group with $N_{C}$ cooperators, $N_{D}$ defectors, and $N-N_{C}-N_{D}$ excluders. These payoffs are given as
\setlength{\arraycolsep}{0.0em}
\begin{eqnarray*}\label{eq2}
\pi_{C}&=&\left\{
\begin{aligned}
&\frac{Fc(N_{C}+N_{E}+1)}{N}(\varsigma-1)+Fc(\langle r\rangle-\varsigma+1)-\langle r\rangle c,  \quad  \text{if} \ \varsigma \leq \langle r\rangle \quad\text{and}\quad N_{E}\neq0;\\
&\frac{Fc(N_{C}+N_{E}+1)}{N}\langle r\rangle-\langle r\rangle c,  \quad  \text{otherwise}.
\end{aligned}
\right.\\
\pi_{D}&=&\left\{
\begin{aligned}
&\frac{Fc(N_{C}+N_{E})}{N}(\varsigma-1),  \quad  \text{if} \ \varsigma \leq \langle r\rangle \quad\text{and}\quad N_{E}\neq0;\\
&\frac{Fc(N_{C}+N_{E})}{N}\langle r\rangle,  \quad  \text{otherwise}.
\end{aligned}
\right.\label{eq3}\\
\pi_{E}&=&\left\{
\begin{aligned}
&\frac{Fc(N_{C}+N_{E}+1)}{N}(\varsigma-1)+Fc(\langle r\rangle-\varsigma+1)-\langle r\rangle c-c_{E}N_{D}-\sigma,     \quad   \text{if} \ \varsigma \leq \langle r\rangle;\\
&\frac{Fc(N_{C}+N_{E}+1)}{N}\langle r\rangle-\langle r\rangle c-\sigma,  \quad  \text{otherwise}.
\end{aligned}\label{function1}
\right.\label{eq4}
\end{eqnarray*}
\setlength{\arraycolsep}{5.0pt}
We first adopt replicator equation to study evolutionary dynamics in infinite well-mixed populations for our model \cite{hofbauer1998evolutionary}. The replicator equation can be written as
\begin{eqnarray}\label{re}
\left\{
\begin{aligned}
\dot{x}&=x(P_{C}-\bar{P}),  \\
\dot{y}&=y(P_{D}-\bar{P}),  \\
\dot{z}&=z(P_{E}-\bar{P}),
\end{aligned}
\right.
\end{eqnarray}
where $\bar{P}=xP_{C}+yP_{D}+zP_{E}$ denotes the average payoff of the whole population. In the following, we investigate the replicator dynamics in detail, including studying the distribution and stability of equilibrium points.

\noindent\subsection{\textbf{Equilibrium points analysis}}

If the average number of rounds is smaller than the number of round in which defectors are excluded, namely, $\langle r\rangle < \varsigma$, then peer exclusion strategy not only degenerates into a pure cooperation strategy, but also needs to pay the additional monitoring cost. Thus defection is a more advantageous strategy than peer exclusion and cooperation.

In the following, we focus on the case of $\varsigma\leq\langle r\rangle$. In this case, the payoffs of cooperators, defectors, and excluders can be respectively written as
\begin{eqnarray*}
\pi_{C}&=&[\frac{Fc(N_{C}+N_{E}+1)}{N}(\varsigma-1)+Fc(\langle r\rangle-\varsigma+1)-\langle r\rangle c]\Theta(N_{E})\\
&+&[1-\Theta(N_{E})][\frac{Fc(N_{C}+1)}{N}\langle r\rangle-\langle r\rangle c],  \\
\pi_{D}&=&\frac{Fc(N_{C}+N_{E})}{N}(\varsigma-1)\Theta(N_{E})+[1-\Theta(N_{E})]\frac{FcN_{C}}{N}\langle r\rangle,  \\
\pi_{E}&=&\frac{Fc(N_{C}+N_{E}+1)}{N}(\varsigma-1)+Fc(\langle r\rangle-\varsigma+1)-\langle r\rangle c-c_{E}N_{D}-\sigma,
\end{eqnarray*}
where $\Theta(N_{E})=1$ if $N_{E}\neq 0$, otherwise $\Theta(N_{E})=0$. Considering that
\begin{eqnarray*}\label{}
&&\sum_{N_{C}=0}^{N-1}\sum_{N_{D}=0}^{N-N_{C}-1}\binom{N-1}{N_{C}}\binom{N-N_{C}-1}{N_{D}}x^{N_{C}}y^{N_{D}}z^{N-N_{C}-N_{D}-1}\frac{Fc(N_{C}+N_{E}+1)}{N}\\
&&=\frac{Fc}{N}[1+(N-1)(1-y)]
\end{eqnarray*}
and
\begin{eqnarray*}\label{}
\sum_{N_{C}=0}^{N-1}\sum_{N_{D}=0}^{N-N_{C}-1}\binom{N-1}{N_{C}}\binom{N-N_{C}-1}{N_{D}}x^{N_{C}}y^{N_{D}}z^{N-N_{C}-N_{D}-1}\langle r\rangle c=\langle r\rangle c,
\end{eqnarray*}
we can accordingly calculate the average payoffs of these three strategies depicted by Eq.~(\ref{eq1}) as
\setlength{\arraycolsep}{0.0em}
\begin{eqnarray*}\label{}
P_{C}&=&\frac{Fc}{N}[1+(N-1)(1-y)](\varsigma-1)+[Fc-(x+y)^{N-1}\frac{Fc(N-1)y}{N(x+y)}](\langle r\rangle-\varsigma+1)-\langle r\rangle c,  \\
P_{D}&=&\frac{Fc}{N}(N-1)(1-y)(\varsigma-1)+(x+y)^{N-1}\frac{Fc(N-1)x}{N(x+y)}(\langle r\rangle-\varsigma+1),  \\
P_{E}&=&\frac{Fc}{N}[1+(N-1)(1-y)](\varsigma-1)+Fc(\langle r\rangle-\varsigma+1)-c_{E}(N-1)y-\langle r\rangle c-\sigma,
\end{eqnarray*}
\setlength{\arraycolsep}{5pt}
where $(N-1)(1-y)$ denotes the expected number of contributors among the $N-1$ co-players.

In the following, we investigate the equilibrium points in the replicator equation described by Eq.~(\ref{re}). Obviously, we know that there exist three vertex equilibrium points, that is, $(x, y, z)=(1, 0, 0)$, $(0, 1, 0)$, and $(0, 0, 1)$. We then explore the interior equilibrium points in the simplex $S_{3}$. For convenience, let us set $\alpha = [\frac{F(\varsigma-1)+NF(\langle r\rangle-\varsigma+1)-N\langle r\rangle}{F(\langle r\rangle-\varsigma+1)(N-1)}]^{\frac{1}{N-1}}$ and $\theta=\frac{\sigma}{[Fc(\varsigma-1)+NcF(\langle r\rangle-\varsigma+1)-Nc\langle r\rangle]/(N\alpha)-c_{E}(N-1)}$.
Solving $P_{C}=P_{D}$ results in $z=1-\alpha$. Similarly, by solving $P_{C}=P_{E}$, we have $y=\theta$. Thus, there is an interior equilibrium point $(\alpha-\theta, \theta, 1-\alpha)$, when $0<\theta<1$, $0<\alpha<1$, and $0<\alpha-\theta<1$.

Then we investigate the boundary equilibrium points on each edge of the simplex $S_{3}$. On the $CD$ edge we have $z=0$, resulting in $\dot{y}=y(1-y)(P_{D}-P_{C})=y(1-y)(1-\frac{F}{N})\langle r\rangle c>0$. Thus the direction of the dynamics goes from $C$ to $D$ and there are no boundary equilibrium points on the $CD$ edge.

On the edge $CE$ we have $y=0$, resulting in $\dot{x}=x(1-x)(P_{C}-P_{E})=x(1-x)\sigma>0$. Thus the direction of the dynamics goes from $E$ to $C$ and there are no boundary equilibrium points on the $CE$ edge.

On the edge $DE$ we have $y + z = 1$, and the equation system becomes $\dot{y}=y(1-y)(P_{D}-P_{E})$. Solving $P_{E}=P_{D}$ results in $y=\frac{Fc(\varsigma-1)/N+Fc(\langle r\rangle-\varsigma+1)-\langle r\rangle c-\sigma}{(N-1)c_{E}}$, which we set as $\xi$. Thus there exists a boundary equilibrium point $(0, \xi, 1-\xi)$ when $0<\xi<1$ is satisfied.

Thus, there are at most five equilibria, namely, $(x, y, z)=(1, 0, 0), (0, 1, 0), (0, 0, 1), (0, \xi, 1-\xi)$, and $(\alpha-\theta, \theta, 1-\alpha)$.

\noindent\subsection{\textbf{The stability of equilibria}}

In the following, we judge the stability of the fixed points in the equation system. Here we set that
\begin{eqnarray}
\left\{
\begin{aligned}\label{system2}
k(x,y)&=x[(1-x)(P_{C}-P_{E})-y(P_{D}-P_{E})],\\
g(x,y)&=y[(1-y)(P_{D}-P_{E})-x(P_{C}-P_{E})].
\end{aligned}
\right.
\end{eqnarray}
Then the Jacobian of the system is
\begin{equation*}
J=\begin{bmatrix}
\frac{\partial{k(x,y)}}{\partial{x}} & \frac{\partial{k(x,y)}}{\partial{y}}\\
\frac{\partial{g(x,y)}}{\partial{x}} & \frac{\partial{g(x,y)}}{\partial{y}}
\end{bmatrix},
\end{equation*}
where
\begin{eqnarray*}
\left\{
\begin{aligned}
\frac{\partial{k(x,y)}}{\partial{x}}&=[(1-x)(P_{C}-P_{E})-y(P_{D}-P_{E})]+x[-(P_{C}-P_{E})+(1-x)\frac{\partial}{\partial{x}}(P_{C}-P_{E})\\
&-y\frac{\partial}{\partial{x}}(P_{D}-P_{E})],\\
\frac{\partial{k(x,y)}}{\partial{y}}&=x[(1-x)\frac{\partial}{\partial{y}}(P_{C}-P_{E})-(P_{D}-P_{E})-y\frac{\partial}{\partial{y}}(P_{D}-P_{E})],\\
\frac{\partial{g(x,y)}}{\partial{x}}&=y[(1-y)\frac{\partial}{\partial{x}}(P_{D}-P_{E})-(P_{C}-P_{E})-x\frac{\partial}{\partial{x}}(P_{C}-P_{E})],\\
\frac{\partial{g(x,y)}}{\partial{y}}&=[(1-y)(P_{D}-P_{E})-x(P_{C}-P_{E})]+y[-(P_{D}-P_{E})+(1-y)\frac{\partial}{\partial{y}}(P_{D}-P_{E})\\
&-x\frac{\partial}{\partial{y}}(P_{C}-P_{E})].
\end{aligned}
\right.
\end{eqnarray*}

The stability analysis of all equilibria is presented below.

$(1)$ For the fixed point $(0, 0, 1)$, the Jacobian is
\begin{equation*}
J=\begin{bmatrix}
\sigma & 0\\
0 & -Fc(\langle r\rangle-\varsigma+1)+\langle r\rangle c+\sigma-\frac{Fc}{N}(\varsigma-1)
\end{bmatrix},
\end{equation*}
thus the fixed equilibrium is unstable since $\sigma>0$.

$(2)$ For the fixed point $(1, 0, 0)$, the Jacobian is
\begin{equation*}
J=\begin{bmatrix}
-\sigma & -\langle r\rangle c+\frac{Fc}{N}\langle r\rangle-\sigma\\
0 & \langle r\rangle c-\frac{Fc}{N}\langle r\rangle
\end{bmatrix},
\end{equation*}
thus the fixed equilibrium is unstable since $F<N$.

$(3)$ For the fixed point $(0, 1, 0)$, the Jacobian is
\begin{equation*}
J=\begin{bmatrix}
\frac{Fc}{N}\langle r\rangle-\langle r\rangle c & 0\\
a_{21} & Fc(\langle r\rangle-\varsigma+1)+\frac{Fc}{N}(\varsigma-1)-c_{E}(N-1)-\langle r\rangle c-\sigma
\end{bmatrix},
\end{equation*}
where $a_{21}=\frac{Fc(N-1)}{N}(\langle r\rangle-\varsigma+1)-c_{E}(N-1)-\sigma$, thus the fixed equilibrium is stable when $Fc(\langle r\rangle-\varsigma+1)+\frac{Fc}{N}(\varsigma-1)-c_{E}(N-1)-\langle r\rangle c-\sigma<0$, while when $Fc(\langle r\rangle-\varsigma+1)+\frac{Fc}{N}(\varsigma-1)-c_{E}(N-1)-\langle r\rangle c-\sigma>0$ it is unstable. Particularly, when $Fc(\langle r\rangle-\varsigma+1)+\frac{Fc}{N}(\varsigma-1)-c_{E}(N-1)-\langle r\rangle c-\sigma=0$, we know that one eigenvalue of the Jacobian is zero and the other eigenvalue is negative. Accordingly, we study its stability by using the center manifold theorem \cite{Khalil1996} in the following.

Because $y = 1 - x - z$, the replicator equation becomes
\begin{eqnarray}\label{system2.2}
\left\{
\begin{aligned}
\dot{x}&=x[(1-x)(P_{C}-P_{D})-z(P_{E}-P_{D})],\\
\dot{z}&=z[(1-z)(P_{E}-P_{D})-x(P_{C}-P_{D})],
\end{aligned}
\right.\label{2.3}
\end{eqnarray}
where
\begin{eqnarray*}
P_{C}-P_{D}&=&\frac{Fc}{N}(\varsigma-1)+Fc(\langle r\rangle-\varsigma+1)-\langle r\rangle c-(1-z)^{N-1}\frac{Fc(N-1)}{N}(\langle r\rangle-\varsigma+1),\\
P_{E}-P_{D}&=&\frac{Fc}{N}(\varsigma-1)+Fc(\langle r\rangle-\varsigma+1)-\langle r\rangle c-\sigma-c_{E}(N-1)(1-x-z)\\
&&-(1-z)^{N-1}\frac{Fc(N-1)x}{N(1-z)}(\langle r\rangle-\varsigma+1).
\end{eqnarray*}

We know that $(x, z)=(0, 0)$ is an equilibrium point of the equation system~(\ref{system2.2}). Then the Jacobian is
\begin{equation*}
A=\begin{bmatrix}
\frac{Fc}{N}\langle r\rangle-\langle r\rangle c & 0\\
0 & Fc(\langle r\rangle-\varsigma+1)+\frac{Fc}{N}(\varsigma-1)-c_{E}(N-1)-\langle r\rangle c-\sigma
\end{bmatrix}.
\end{equation*}
When $Fc(\langle r\rangle-\varsigma+1)+\frac{Fc}{N}(\varsigma-1)-c_{E}(N-1)-\langle r\rangle c-\sigma=0$, we know that the eigenvalues of the Jacobi matrix for the fixed point $(x, z)=(0, 0)$ are $\frac{Fc}{N}\langle r\rangle-\langle r\rangle c$ and  $0$. In this condition, we have that $x = h(z)$ is a center manifold of the equation system. We start to try $h(z)=O(z^2)$, thus the equation system can be expressed as
\begin{eqnarray*}
\dot{z}&=&z(1-z)[\frac{Fc}{N}(\varsigma-1)+Fc(\langle r\rangle-\varsigma+1)-\langle r\rangle c-\sigma-c_{E}(N-1)(1-z)]\\
&=&z^{2}(1-z)(N-1)c_{E}+O(|z|^{4}).
\end{eqnarray*}
Since $c_{E}(N-1)\neq0$, thus we can judge that $z=0$ is unstable. Accordingly, the fixed point $(0, 1, 0)$ is unstable as well.

$(4)$ For $(x, y, z)=(0, \xi, 1-\xi)$, the Jacobian is
\begin{equation*}
J=\begin{bmatrix}
a_{11} & 0\\
a^{*}_{21} & a_{22}
\end{bmatrix},
\end{equation*}
where $a_{11}=c_{E}(N-1)\xi+\sigma-\xi^{N-1}\frac{Fc(N-1)}{N}(\langle r\rangle-\varsigma+1), a^{*}_{21}=\xi^{N-1}\frac{Fc(N-1)(\langle r\rangle-\varsigma+1)}{N}-c_{E}(N-1)\xi^{2}-\sigma \xi$, and $a_{22}=\xi(1-\xi)(N-1)c_{E}$, thus the fixed equilibrium is unstable since $\xi(1-\xi)(N-1)c_{E}>0$.

$(5)$ For $(x,y,z)=(\alpha-\theta, \theta, 1-\alpha)$, we define the equilibrium point as $(x^{*},y^{*},z^{*})$ hereafter, thus the elements in the Jacobian are written as
\begin{eqnarray*}
\left\{
\begin{aligned}
\frac{\partial{k}}{\partial{x}}(x^{*},y^{*})&=&x^{*}[(1-x^{*})\frac{\partial}{\partial{x}}(P_{C}-P_{E})-y^{*}\frac{\partial}{\partial{x}}(P_{D}-P_{E})],\\
\frac{\partial{k}}{\partial{y}}(x^{*},y^{*})&=&x^{*}[(1-x^{*})\frac{\partial}{\partial{y}}(P_{C}-P_{E})-y^{*}\frac{\partial}{\partial{y}}(P_{D}-P_{E})],\\
\frac{\partial{g}}{\partial{x}}(x^{*},y^{*})&=&y^{*}[(1-y^{*})\frac{\partial}{\partial{x}}(P_{D}-P_{E})-x^{*}\frac{\partial}{\partial{x}}(P_{C}-P_{E})],\\
\frac{\partial{g}}{\partial{y}}(x^{*},y^{*})&=&y^{*}[(1-y^{*})\frac{\partial}{\partial{y}}(P_{D}-P_{E})-x^{*}\frac{\partial}{\partial{y}}(P_{C}-P_{E})],
\end{aligned}
\right.
\end{eqnarray*}
where
\begin{eqnarray*}
\left\{
\begin{aligned}
\frac{\partial}{\partial{x}}(P_{C}-P_{E})&=-(N-2)(x+y)^{N-3}\frac{Fc(N-1)y}{N}(\langle r\rangle-\varsigma+1),\\
\frac{\partial}{\partial{y}}(P_{C}-P_{E})&=c_{E}(N-1)-(x+y)^{N-3}\frac{Fc(N-1)}{N}(\langle r\rangle-\varsigma+1)[(N-1)y+x],\\
\frac{\partial}{\partial{x}}(P_{D}-P_{E})&=(x+y)^{N-3}\frac{Fc(N-1)}{N}(\langle r\rangle-\varsigma+1)[(N-1)x+y],\\
\frac{\partial}{\partial{y}}(P_{D}-P_{E})&=c_{E}(N-1)+(x+y)^{N-3}\frac{Fc(N-1)x}{N}(\langle r\rangle-\varsigma+1)(N-2).
\end{aligned}
\right.
\end{eqnarray*}
Then we define that $p^{*}=\frac{\partial{k}}{\partial{x}}(x^{*},y^{*})\frac{\partial{g}}{\partial{y}}(x^{*},y^{*})-\frac{\partial{k}}{\partial{y}}(x^{*},y^{*})\frac{\partial{g}}{\partial{x}}(x^{*},y^{*})$ and $q^{*}=\frac{\partial{k}}{\partial{x}}(x^{*},y^{*})+\frac{\partial{g}}{\partial{y}}(x^{*},y^{*})$. Thus we have
\begin{eqnarray*}
p^{*}&=&\frac{\partial{k}}{\partial{x}}(x^{*},y^{*})\frac{\partial{g}}{\partial{y}}(x^{*},y^{*})-\frac{\partial{k}}{\partial{y}}(x^{*},y^{*})\frac{\partial{g}}{\partial{x}}(x^{*},y^{*})\nonumber\\
&=&x^{*}y^{*}(1-x^{*}-y^{*})[\frac{\partial}{\partial{x}}(P_{C}-P_{E})\frac{\partial}{\partial{y}}(P_{D}-P_{E})-\frac{\partial}{\partial{y}}(P_{C}-P_{E})\frac{\partial}{\partial{x}}(P_{D}-P_{E})]\nonumber\\
&=&x^{*}y^{*}(1-x^{*}-y^{*})\big\{\frac{Fc(\langle r\rangle-\varsigma+1)}{N}(N-1)^{3}(x^{*}+y^{*})^{N-2}[(x^{*}+y^{*})^{N-2}\frac{Fc}{N}(\langle r\rangle\nonumber\\
&-&\varsigma+1)-c_{E}]\big\}
\end{eqnarray*}
and
\begin{eqnarray*}
q^{*}&=&\frac{\partial{k}}{\partial{x}}(x^{*},y^{*})+\frac{\partial{g}}{\partial{y}}(x^{*},y^{*})\nonumber\\
&=&x^{*}(1-x^{*})\frac{\partial}{\partial{x}}(P_{C}-P_{E})+y^{*}(1-y^{*})\frac{\partial}{\partial{y}}(P_{D}-P_{E})-x^{*}y^{*}[\frac{\partial}{\partial{x}}(P_{D}-P_{E})+\frac{\partial}{\partial{y}}(P_{C}-P_{E})]\nonumber\\
&=&(N-1)c_{E}y^{*}(1-x^{*}-y^{*})\nonumber\\
&>&0.
\end{eqnarray*}
Therefore the interior equilibrium point is unstable since the largest eigenvalue of Jacobian is positive.

\noindent\subsection{\textbf{Heteroclinic cycle}}

In this subsection, we show that there exists a stable heteroclinic cycle on the boundary of the simplex $CDE$. When $\varsigma < \frac{N[(F-1)\langle r\rangle c-\sigma-(N-1)c_{E}]}{(N-1)Fc}+1$ and $F<N$, we know that the three vertex equilibrium points ($C, D$, and $E$) are all saddle nodes, and the heteroclinic trajectories can display on the three edges ($CD, DE$, and $EC$). All of these guarantee the existence of the heteroclinic cycle on the boundary of the simplex $S_{3}$. Based on previous works~\cite{Han2003,Afraimovich2004,Shaw2012,Park2018}, we give an applicable theorem to prove that the heteroclinic cycle is asymptotically stable as follows.

\noindent \textbf{Theorem 1}
Consider a planar system
\begin{eqnarray}\label{hc}
\left\{
\begin{aligned}
\dot{x_{1}}&=f_{1}(x_{1},x_{2}),  \\
\dot{x_{2}}&=f_{2}(x_{1},x_{2}),  \\
\end{aligned}
\right.
\end{eqnarray}
where $f_{1}$ and $f_{2}$ are continuous functions on the plane. Suppose that system (\ref{hc}) has a heteroclinic cycle
$L=L_{1}\bigcup L_{2}\bigcup L_{3}$ with three different saddle points $\Psi_{i}$, $i=1,2,3$. Let $\lambda_{\Psi_{i}1}>0$ and $\lambda_{\Psi_{i}2}<0$
be the eigenvalues of the Jacobian $J(\Psi_{i})$. Then the hyperbolicity ratio of $\Psi_{i}$ is given by $\lambda_{i}=-\frac{\lambda_{\Psi_{i}2}}{\lambda_{\Psi_{i}1}}$. Thus, \\
(1) if $\prod_{i=1}^{3}\lambda_{i}>1$, then $L$ is stable;\\
(2) if $\prod_{i=1}^{3}\lambda_{i}<1$, then $L$ is unstable.

The above theorem has been proved and widely used in some theoretical studies \cite{Han2003,Afraimovich2004,Shaw2012,Park2018}. In the following, we will apply the above theorem to prove that the heteroclinic cycle is asymptotically stable. According to stability analysis of equilibria, we can get the eigenvalues of the Jacobian of the three vertex equilibrium points as
\begin{eqnarray}
\left\{
\begin{aligned}
\lambda_{E}^{-}&=-Fc(\langle r\rangle-\varsigma+1) + \langle r\rangle c + \sigma-\frac{Fc}{N}(\varsigma-1), \\
\lambda_{E}^{+}&=\sigma, \\
\lambda_{C}^{-}&=-\sigma, \\
\lambda_{C}^{+}&=\langle r\rangle c-\frac{Fc}{N}\langle r\rangle, \\
\lambda_{D}^{-}&=\frac{Fc}{N}\langle r\rangle-\langle r\rangle c, \\
\lambda_{D}^{+}&=Fc(\langle r\rangle-\varsigma+1) - \langle r\rangle c - \sigma+\frac{Fc}{N}(\varsigma-1)-(N-1)c_{E}.
\end{aligned}
\right.
\end{eqnarray}
Then we define that $\lambda_{E}=-\frac{\lambda_{E}^{-}}{\lambda_{E}^{+}}, \lambda_{C}=-\frac{\lambda_{C}^{-}}{\lambda_{C}^{+}},$ and $\lambda_{D}=-\frac{\lambda_{D}^{-}}{\lambda_{D}^{+}}$, and we have $\lambda=\lambda_{E}\lambda_{C}\lambda_{D}=\frac{Fc(\langle r\rangle-\varsigma+1) - \langle r\rangle c - \sigma+Fc(\varsigma-1)/N}{Fc(\langle r\rangle-\varsigma+1) - \langle r\rangle c - \sigma+Fc(\varsigma-1)/N-(N-1)c_{E}}>1$.
Thus we can judge that the heteroclinic cycle is asymptotically stable according to \textbf{Theorem 1}.

\begin{figure*}
\centering
\includegraphics[width=\textwidth]{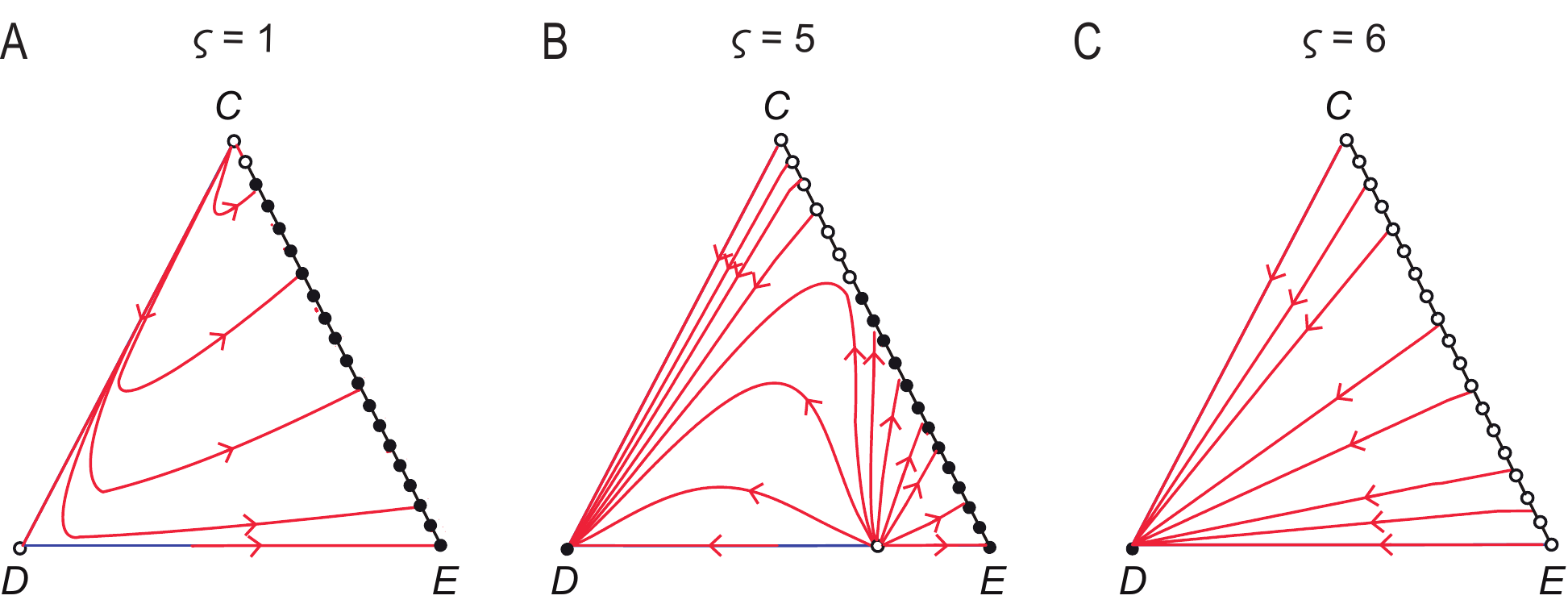}
\begin{flushleft}
\textbf{Supplementary Figure 1: Replicator dynamics of cooperators, defectors, and excluders for different values of exclusion round $\varsigma$ when the monitoring cost $\sigma$ is zero.} The edge $EC$ consists of a continuum of equilibria. For small $\varsigma$, $EC$ is separated into a stable segment and an unstable segment. When $\varsigma$ reaches a critical value, there is an unstable equilibrium point on the $ED$ edge, and the system can reach to one of stable equilibrium points-either all defector state (vertex $D$) or the coexistence state of cooperators and excluders. For even larger $\varsigma (\varsigma>\langle r\rangle)$, all trajectories converge to a state where there are almost $D$ players in the population. Here, $\varsigma=1$ in panel A, $\varsigma=5$ in panel B, and $\varsigma=6$ in panel C. Other parameters values are $N=5$, $F=3$, $c=1$, $c_{E}=0.4, \sigma=0$, and $w=0.8$.
\end{flushleft}
\label{figs1}
\end{figure*}

In Fig.~S1 we investigate the replicator dynamics of cooperators, defectors, and excluders for three different values of exclusion round $\varsigma$ when the monitoring cost $\sigma$ is zero. When $\varsigma$ is small (e.g., $\varsigma=1$), we can find that the whole edge $CE$ consists of a continuum of equilibrium points, and is separated into one stable segment and one unstable segment. The evolution on the edge $CD$ is unidirectional from $C$ to $D$, and from $D$ to $E$ on the edge $DE$. All interior trajectories will converge to the stable segment of $CE$ edge (Fig.~S1A), which recovers previous results that excluders can emerge in a sea of defectors and dominate them \cite{sasaki2013evolution}. As $\varsigma$ increases, an unstable equilibrium appears on the $ED$ edge. The interior space is occupied by the basins of attraction of $D$ and stable segment on the $CE$ edge (Fig.~S1B). As $\varsigma$ continues to increase and exceeds the average number of round $\langle r\rangle$, defectors can occupy the whole population (Fig.~S1C).

\begin{figure*}
\centering
\includegraphics[width=\textwidth]{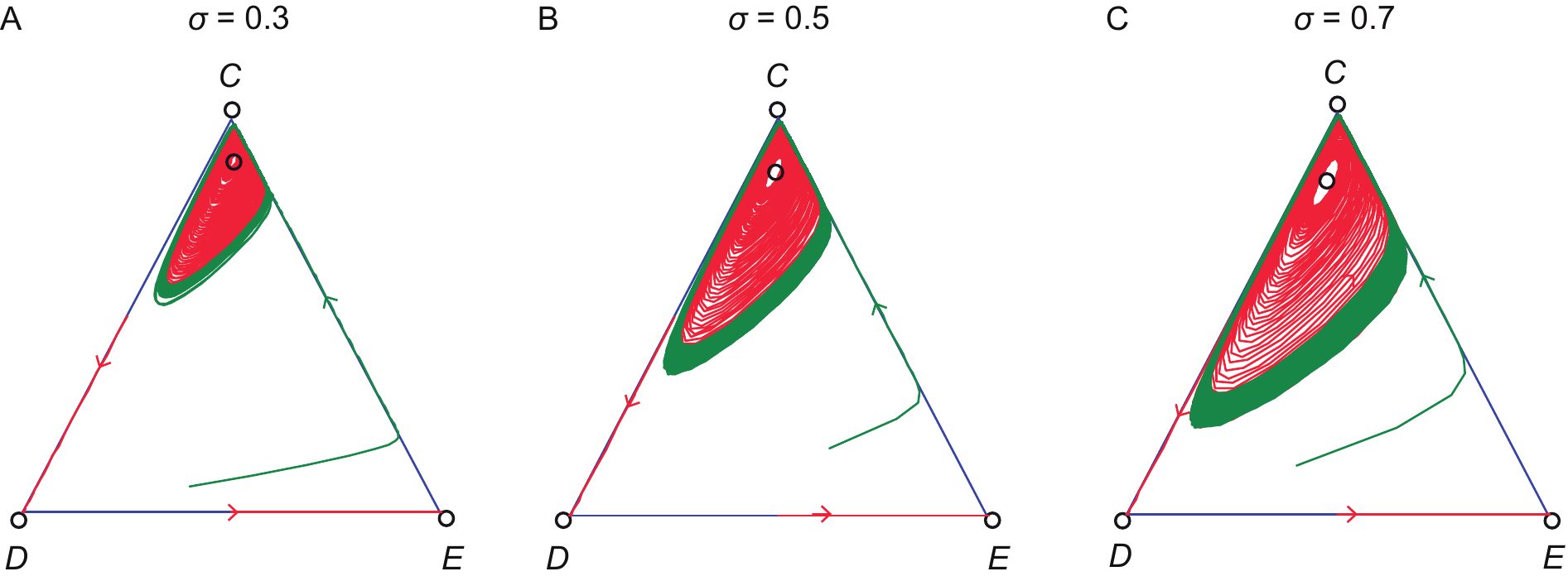}
\begin{flushleft}
\textbf{Supplementary Figure 2: Replicator dynamics of cooperators, defectors, and excluders in an infinite population for three different monitoring cost $\sigma$ when the exclusion round $\varsigma$ is small.} The starting points are different on the curves with different colors. The system will converge to the limit cycle in the interior of simplex when the monitoring cost is changed appropriately. Here, $\sigma=0.3$ in panel A, $\sigma=0.5$ in panel B, and $\sigma=0.7$ in panel C. Other parameters values are $N=5$, $F=3$, $c=1$, $c_{E}=0.4, \varsigma=2$, and $w=0.8$.\label{figs2}
\end{flushleft}
\end{figure*}

Furthermore, we investigate the effect of different values of monitoring cost on the replicator dynamics. As shown in Fig.~S2, we show that our previous results remain valid if the value of monitoring cost is approximately changed, namely, cooperation can be maintained in the population by forming evolutionary oscillations with other two strategies. Furthermore, we find that the oscillations dynamics may be affected when the model parameters are changed significantly. As shown in Fig.~S3, when the value of exclusion cost $c_{E}$ ranges from 0.3 to 1, for low $\sigma$ the coexistence state of $C$, $D$, and $E$ can be dominant. While when $\sigma$ values further increase and exceed a certain threshold, defection is the dominant strategy.

\begin{figure*}
\centering
\includegraphics[width=.7\linewidth]{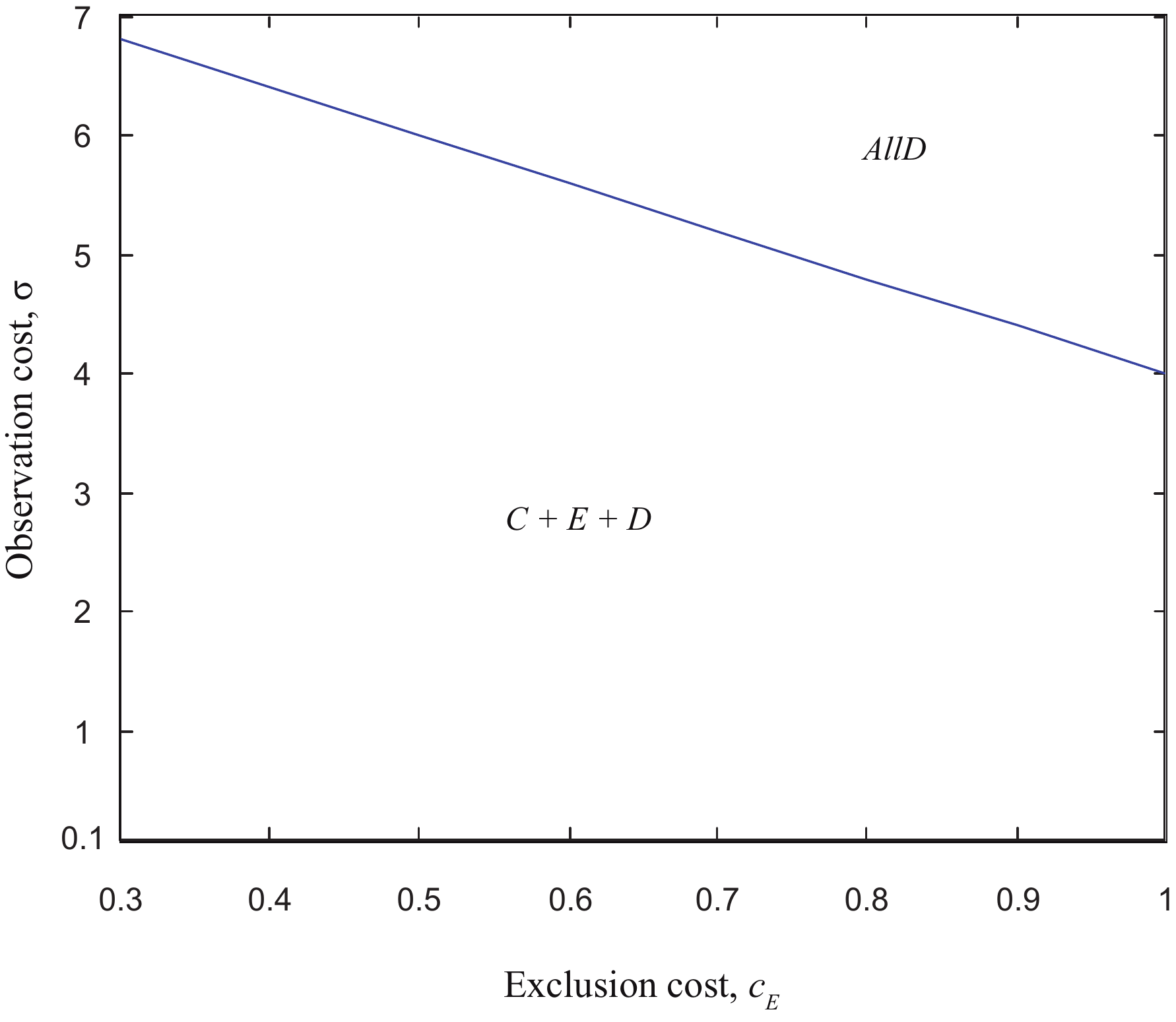}
\begin{flushleft}
\textbf{Supplementary Figure 3: Full $c_{E}$-$\sigma$ phase diagrams.} Cooperators ($C$), defectors ($D$), and excluders ($E$) can form cyclic dominant dynamics when the observation cost $\sigma$ is relatively small. For a given $c_{E}$, when $\sigma$ exceeds a certain threshold, defectors can occupy the whole population. Parameters are $N=5$, $F=3$, $c=1$, $\varsigma=6$, and $w=0.9$.\label{phase}
\end{flushleft}
\end{figure*}

\begin{figure*}
\centering
\includegraphics[width=.9\linewidth]{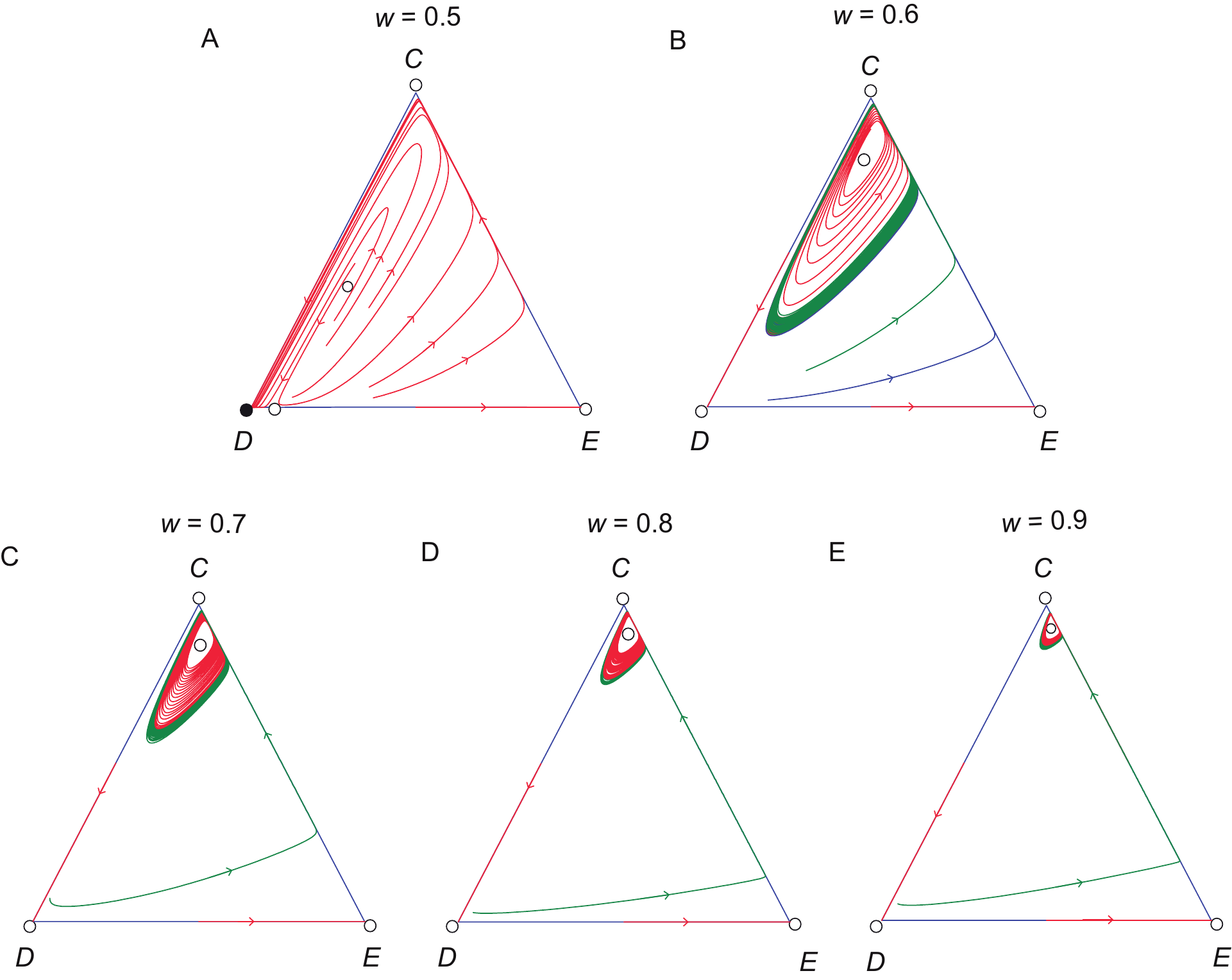}
\begin{flushleft}
\textbf{Supplementary Figure 4: Replicator dynamics of cooperators, defectors, and excluders in an infinite population for different values of discount factor $w$.} The starting points are different on the curves with different colors. Defection is the most advantageous strategy when $w$ takes an intermediate value. A further increase in $w$ leads the system to converging to a limit cycle in the interior of the simplex. Here, $w=0.5$ in panel A, $w=0.6$ in panel B, $w=0.7$ in panel C, $w=0.8$ in panel D, and $w=0.9$ in panel E. Other parameters values are $N=5$, $F=3$, $c=1$, $c_{E}=0.4, \varsigma=2$, and $\sigma=0.1$.
\end{flushleft}
\label{figs3}
\end{figure*}

We also show the replicator dynamics for different discounted factor $w$ when exclusion round $\varsigma$ is fixed at $2$, as shown in Fig.~S4. For intermediate $w$, cooperation cannot be sustained at all because excluding free-riders in the last round cannot effectively weaken their overall advantage (Fig.~S4A). As the discounted factor $w$ increases, the advantage of defectors is weakened. Hence excluders can do better than defectors when there are no cooperators in the group, which leads to that the evolutionary direction is from $D$ to $E$ on the boundary $DE$. Besides, cooperators can do better than excluders on the boundary $CE$, while they have fewer advantages than defectors on the boundary $CD$. Furthermore, the interior trajectories will form the evolutionary oscillations in the interior of the simplex, which implies that the three strategies can coexist in the population (Fig.~S4B). For even larger $w$, numerical results show that the amplitude of oscillations is decreasing gradually (Fig.~S4B-E).

\noindent\subsection{\textbf{Replicator-mutator dynamics}}

\begin{figure*}
\centering
\includegraphics[width=\textwidth]{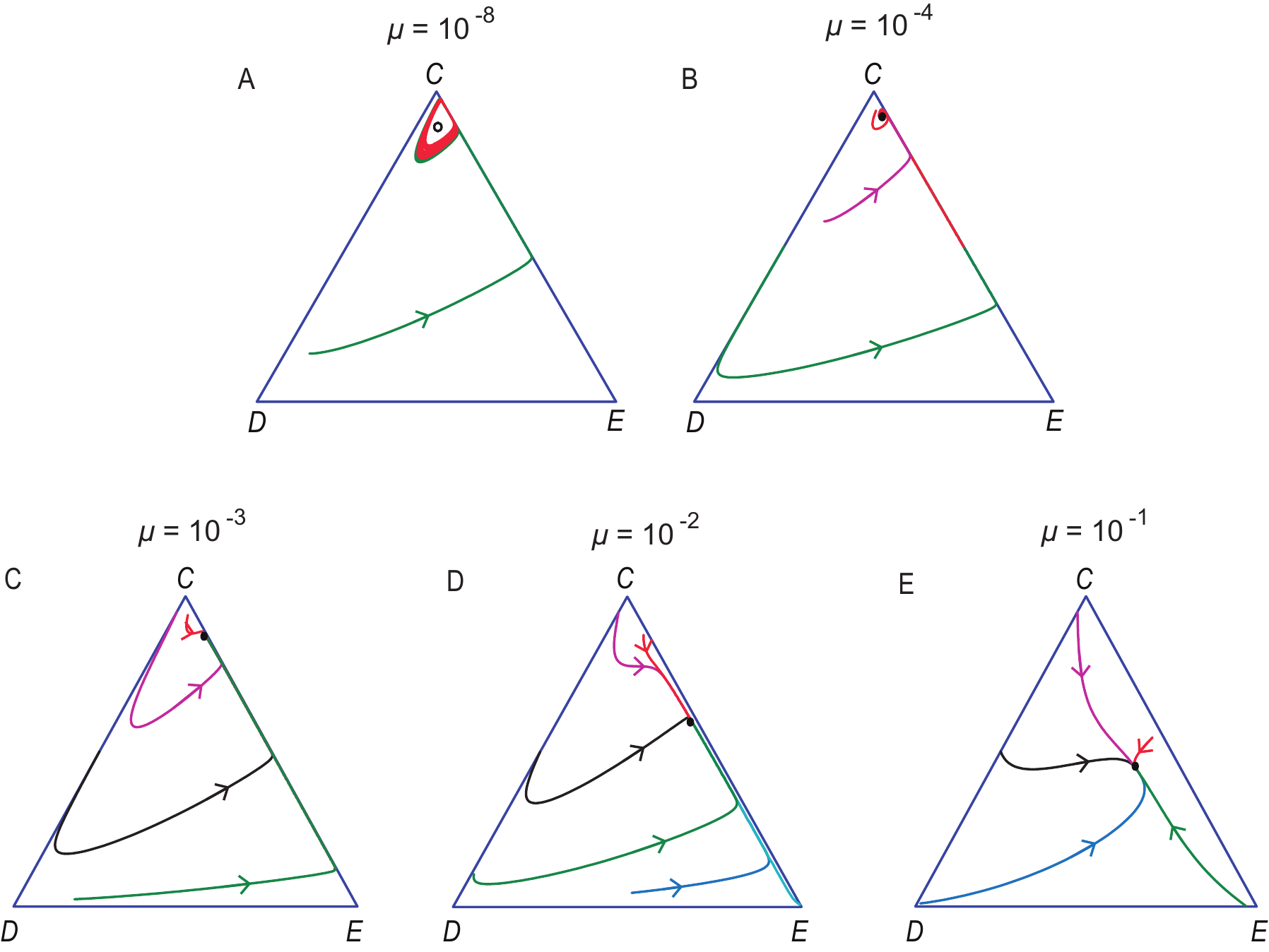}
\begin{flushleft}
\textbf{Supplementary Figure 5: Replicator-mutator dynamics of cooperators, defectors, and excluders in an infinite population for different mutation rates $\mu$ when the value of exclusion round is small.} The starting points are different on the curves with different colors. For very small mutation rates, there is an interior equilibrium point which is unstable. Besides, evolutionary oscillations occur around this equilibrium point. When the mutation rate exceeds a critical value, the oscillations disappear and all trajectories converge to a stable fixed point at which all three strategies coexist. Here, $\mu=10^{-8}$ in panel A, $\mu=10^{-4}$ in panel B, $\mu=10^{-3}$ in panel C, $\mu=10^{-2}$ in panel D, and $\mu=10^{-1}$ in panel E. Other parameters values are $N=5$, $F=3$, $c=1$, $c_{E}=0.4, \sigma=0.1$, $w=0.8$, and $\varsigma=2$.\label{figs4}
\end{flushleft}
\end{figure*}

Here we introduce mutation into the replicator equation \cite{page2002unifying} and the ``replicator-mutator" equation is written as
\begin{eqnarray}
\left\{
\begin{aligned}
\dot{x}&=xP_{C}q_{C\rightarrow C}+yP_{D}q_{D\rightarrow C}+zP_{E}q_{E\rightarrow C}-x\bar{P},\\
\dot{y}&=yP_{D}q_{D\rightarrow D}+xf_{C}q_{C\rightarrow D}+zP_{E}q_{E\rightarrow D}-y\bar{P},\\
\dot{z}&=zP_{E}q_{E\rightarrow E}+xP_{C}q_{C\rightarrow E}+yP_{D}q_{D\rightarrow E}-z\bar{P},
\end{aligned}
\right.
\end{eqnarray}
where $q_{U\rightarrow V}$ denotes the probability that an offspring of $U$ strategist adopts strategy $V$. Following previous work \cite{imhof2005evolutionary}, we set that $q_{U\rightarrow V}=\mu$ for $U\neq V$ and $q_{U\rightarrow U}=1-2\mu$.

In the resulting ``replicator-mutator" equation, the evolutionary dynamics remain essentially the same for very small mutation rates: there is an interior equilibrium point which is unstable and all trajectories starting in the interior space form the evolutionary oscillations around this interior equilibrium point (Fig.~S5A). When the mutation rate exceeds a critical value, the oscillations disappear and all trajectories converge to a stable interior fixed point (Fig.~S5B). Besides, as the mutation rate $\mu$ increases, the stable equilibrium point moves towards the center of simplex gradually (Fig.~S5C-E).\\

\noindent\section{\textbf{Evolutionary dynamics in finite well-mixed populations}}
\renewcommand\thesubsection{2.\arabic{subsection}}

Considering that in real society, the size of population is usually finite, and the deterministic equations in last section are not suitable for such population sizes, because stochastic effects and random drift play important roles. In finite well-mixed populations of size $Z$ with $i_{C}$ cooperators, $i_{E}$ excluders, and $Z-i_{C}-i_{E}$ defectors, the average payoffs of cooperators, defectors, and excluders can be, respectively, computed by using a multivariate hypergeometric sampling, as
\begin{eqnarray}\label{sec21}
f_{C}(i_{E},i_{C})&=&\sum_{N_{C}=0}^{N-1}\sum_{N_{E}=0}^{N-N_{C}-1}\frac{\binom{i_{C}-1}{N_{C}}\binom{i_{E}}{N_{E}}\binom{Z-i_{C}-i_{E}}{N-N_{C}-N_{E}-1}}{\binom{Z-1}{N-1}}\pi_{C},\nonumber\\
f_{D}(i_{E},i_{C})&=&\sum_{N_{C}=0}^{N-1}\sum_{N_{E}=0}^{N-N_{C}-1}\frac{\binom{i_{C}}{N_{C}}\binom{i_{E}}{N_{E}}\binom{Z-i_{C}-i_{E}-1}{N-N_{C}-N_{E}-1}}{\binom{Z-1}{N-1}}\pi_{D},\\
f_{E}(i_{E},i_{C})&=&\sum_{N_{C}=0}^{N-1}\sum_{N_{E}=0}^{N-N_{C}-1}\frac{\binom{i_{C}}{N_{C}}\binom{i_{E}-1}{N_{E}}\binom{Z-i_{C}-i_{E}}{N-N_{C}-N_{E}-1}}{\binom{Z-1}{N-1}}\pi_{E}.\nonumber
\end{eqnarray}

We then adopt the pairwise comparison rule to describe the evolutionary dynamics of $C, E$, and $D$ in a finite population. At each discrete time step, one player $L$ randomly selected from the population updates his/her strategy. We consider the possibility of mutation, hence player $L$ adopts a randomly chosen available strategy with probability $\mu$. Alternatively, with probability $1-\mu$, player $L$ tends to adopt the strategy of another randomly chosen player $R$ with a probability given by the Fermi function
\begin{eqnarray*}
\frac{1}{1+\exp{[\beta(f_{L}-f_{R})}]},
\end{eqnarray*}
where $\beta$ denotes the intensity of selection~\cite{szabo_pre98}. In the $\beta \rightarrow \infty$ strong selection limit, the more successful player always succeeds in enforcing his/her strategy to player $L$, but never otherwise. $\beta \rightarrow 0$ indicates the so-called weak selection limit where strategy adoption becomes random independently of the average payoff difference. In between these extremes, for a finite value of $\beta$, it is likely that a better performing strategy is imitated, but it is still not impossible to adopt his/her strategy when performing worse.

In the following, we focus on the case of $\varsigma\leq\langle r\rangle$. We first numerically investigate the stochastic dynamics among cooperators, defectors, and excluders when the mutation rate is arbitrarily large. Then we present the theoretical analysis of the evolutionary dynamics of the system when mutation rate is sufficiently small.

\noindent\subsection{\textbf{Arbitrarily mutation rates}}

For an arbitrarily mutation rate, the time scales between imitation and mutation are no longer separated, and the population state is not homogeneous most of the time. In this situation, all strategies are always present in the population state. Besides, the evolutionary dynamics of the system are also affected by noise arising from the finite population. All of these render the stationary distribution as the most appropriate quantity to analyse the behaviour of the population.

To describe the stationary distribution of these three strategies in finite populations, we adopt the following Master-Equation method. As the evolution of the system depends only on its actual configuration, evolutionary dynamics of cooperation, defection, and exclusion can be described as a Markov process over a two-dimensional space. The study of Markov process consists in determining its probability density function evolution, $p_{\textbf{i}}(t)$, which provides information on the prevalence of each configuration at time $t$. Since $\textbf{i}(t)$ has the Markov property, its transition probability and probability density function obey the discrete time Master-Equation~\cite{kampen2007}. The mentioned Master-Equation is written as
\begin{eqnarray}\label{eq14}
p_{\textbf{i}}(t+\tau)-p_{\textbf{i}}(t)=\sum_{\textbf{i}^{'}}\{T_{\textbf{i}\textbf{i}^{'}}p_{\textbf{i}^{'}}(t)-T_{\textbf{i}^{'}\textbf{i}}p_{\textbf{i}}(t)\},
\end{eqnarray}
where $T_{\textbf{i}\textbf{i}^{'}}$ represents the transition probability from configuration $\textbf{i}^{'}$ to configuration $\textbf{i}$, and the transition from state $\textbf{i}$ to state $\textbf{i}^{'}$ is denoted by $T_{\textbf{i}^{'}\textbf{i}}$. Figure~S6 illustrates a concrete example of a local phase space and all six possible transitions in one-step process. Then the stationary distribution $\bar{p_{\textbf{i}}}$ can be obtained by making the left-hand side of Eq.~(\ref{eq14}) be equal to zero. Technically, we can search the eigenvector corresponding to the eigenvalue 1 of the transition matrix to get the stationary distribution of the system.

\begin{figure*}
\centering
\includegraphics[width=9cm]{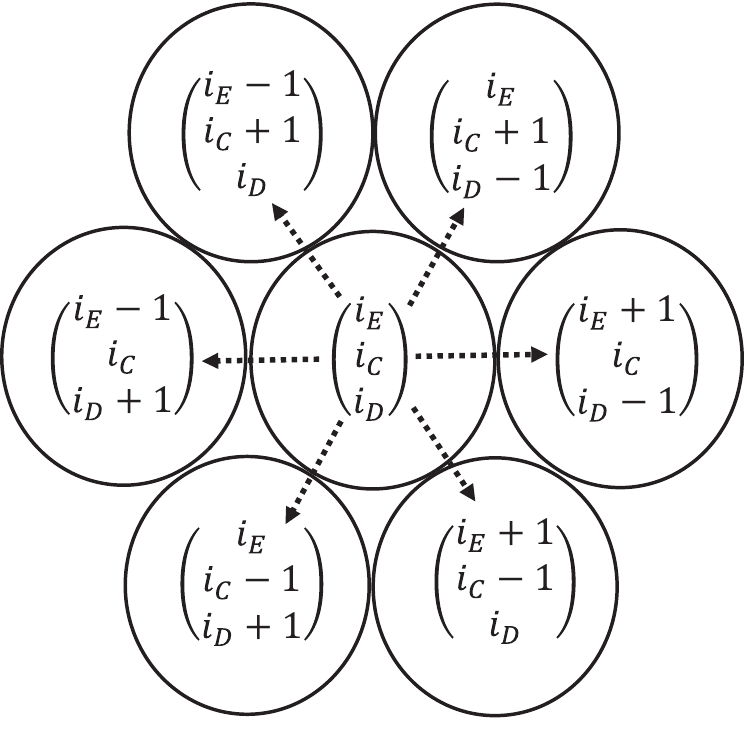}
\begin{flushleft}
\textbf{Supplementary Figure 6: The representation of local phase space and possible transitions in one-step process.} The vector in the central circle represents the configuration of the population at a given time $t$. The vectors of the surrounding six circles represent the adjacent six configurations that may be reached at the next time step \cite{Vasconcelos2013A}.
\end{flushleft}
\label{figS5}
\end{figure*}

Now, we turn our attention to computing the transition probabilities $T_{\textbf{i}\textbf{i}^{'}}$. Considering a finite population with $Z$ players who can opt for the strategies $C, D$, or $E$. Let $\textbf{i}(t)=\{i_{E}, i_{C}\}$ denote the possible configuration of strategies at a given time. At the next time step, it can either move to one of the adjacent configurations $\textbf{i}^{'}(t+1)=\{i_{E}^{'},i_{C}^{'}\}$ (see Fig.~S6) or remain unchanged. The transition probability between two adjacent configurations corresponds to the probability that one player with strategy $U$ adopts another different strategy $V$ at the next time step. We thus have
\begin{eqnarray}\label{eq15}
T_{U\rightarrow V}=(1-\mu)\big[\frac{i_{U}}{Z}\frac{i_{V}}{Z-1}\frac{1}{1+\exp({\beta(f_{U}-f_{V}))}}\big]+\mu\frac{i_{U}}{(d-1)Z},
\end{eqnarray}
where $U, V = C, D,$ or $E$, and $d$ denotes the number of strategies in the strategy space, here we have $d=3$. Besides, when the system remains unchanged, the transition probability can be calculated as $T_{\textbf{i}\textbf{i}}=1-\sum_{\textbf{i}^{'}\neq \textbf{i}}T_{\textbf{i}^{'}\textbf{i}}$.

In addition to the analysis of the stationary distribution, another quantity of interest in studying the evolutionary dynamics in finite populations is the gradient of selection, which describes the most possible evolutionary direction in phase space of three different strategies. Following previous work \cite{imhof2005evolutionary}, we can employ a Kramers-Moyal expansion of the Master Equation~(\ref{eq14}) to derive a Fokker-Planck equation where the first coefficient corresponds to the gradient of selection. Consequently, for a given configuration $\textbf{i}=\{i_{E},i_{C}\}$, we can compute the gradient of selection as
\begin{eqnarray}
g_{\textbf{i}}=(T_{\textbf{i}}^{E+}-T_{\textbf{i}}^{E-})\textbf{u}_{1}+(T_{\textbf{i}}^{C+}-T_{\textbf{i}}^{C-})\textbf{u}_{2},
\end{eqnarray}
where $\textbf{u}_{1}$ and $\textbf{u}_{2}$ are a set of unit vectors with the direction of $\frac{\partial\textbf{i}}{\partial i_{E}}$ and $\frac{\partial\textbf{i}}{\partial i_{C}}$ respectively. Besides, $T_{\textbf{i}}^{E\pm}$ represents the probability to increase (decrease) by one the number of individuals adopting strategy $E$, and $T_{\textbf{i}}^{C\pm}$ represents the probability to increase (decrease) by one the number of individuals adopting strategy $C$, which can be respectively computed as
\begin{eqnarray*}
T_{\textbf{i}}^{E\pm}=T_{\textbf{i}\{i_{E}\pm1,i_{C}\mp1,i_{D}\}}+T_{\textbf{i}\{i_{E}\pm1,i_{C},i_{D}\mp1\}}
\end{eqnarray*}
and
\begin{eqnarray*}
T_{\textbf{i}}^{C\pm}=T_{\textbf{i}\{i_{E}\mp1,i_{C}\pm1,i_{D}\}}+T_{\textbf{i}\{i_{E},i_{C}\pm1,i_{D}\mp1\}},
\end{eqnarray*}
where the terms on the right side can be obtained by Eq.~(\ref{eq15}).

Finally, we use a key quantity $\bar{\rho}_{U}$ to describe the average level of strategy $U$. It can be computed by averaging over all possible configurations $\textbf{i}$, each weighted with the corresponding stationary distribution $\bar{p_{\textbf{i}}}$, which is given as
\begin{eqnarray}
\bar{\rho}_{U}=\sum_{\textbf{i}}\textbf{i}_{i_{U}}\bar{p_{\textbf{i}}}/Z,
\end{eqnarray}
where $U=C, D,$ or $E$, and $\textbf{i}_{i_{U}}$ denotes the number of strategy $U$ in configuration $\textbf{i}$.

\begin{figure*}
\centering
\includegraphics[width=\textwidth]{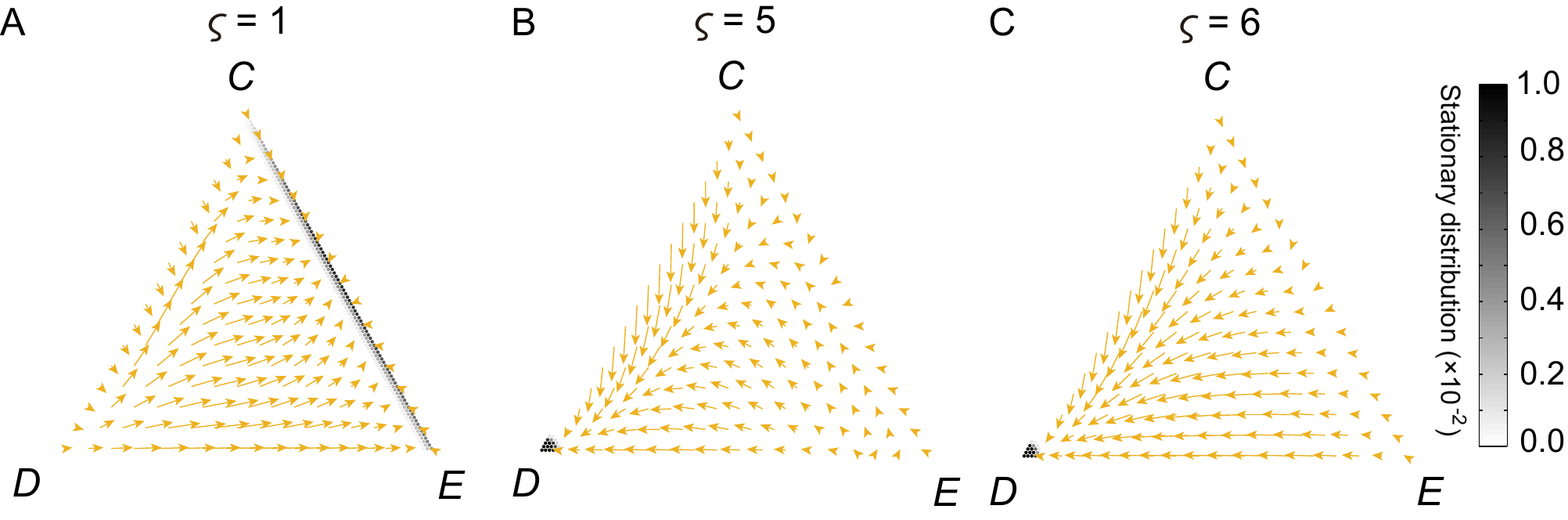}
\begin{flushleft}
\textbf{Supplementary Figure 7: Evolutionary dynamics of cooperators, defectors, and excluders in a finite population for different values of exclusion round $\varsigma$ when the monitoring cost $\sigma$ is zero.} Panel A shows that the population spends most of the time on configurations where $C$ and $E$ players coexist, while panels B and C show that the population spends most of time in configurations which are close to the state of all defection. Here, $\varsigma=1$ in panel A, $\varsigma=5$ in panel B, and $\varsigma=6$ in panel C. Other parameters values are $Z=100, N=5$, $F=3$, $c=1$, $c_{E}=0.4, \beta=2, \mu=10^{-2}$, $w=0.8$, and $\sigma=0$.
\end{flushleft}
\label{figs6}
\end{figure*}

In what follows, we provide numerical results for high mutation rates. We are first interested in studying how these three strategies evolve without considering the cost of monitoring. When exclusion round $\varsigma$ is small (e.g., $\varsigma=1$), defectors rapidly outcompete cooperators due to the social dilemma (see arrows), leading the population to moving towards to the state of full defection. In this situation, however, defectors will be outcompeted by peer excluders, which leads the population to moving towards the state of full cooperation. Since the monitoring cost is zero, the second-order free-riding dilemma disappears, the population spends most time in configurations around the $CE$-edge of the simplex. All these results are summarized in Fig.~S7A. For large exclusion round $\varsigma$, however, the situation is quite different, as shown in Fig.~S7B and C, where excluders and cooperators cannot resist the invasion of defectors, leading the population to moving towards the state of full defection.

\begin{figure*}
\centering
\includegraphics[width=\textwidth]{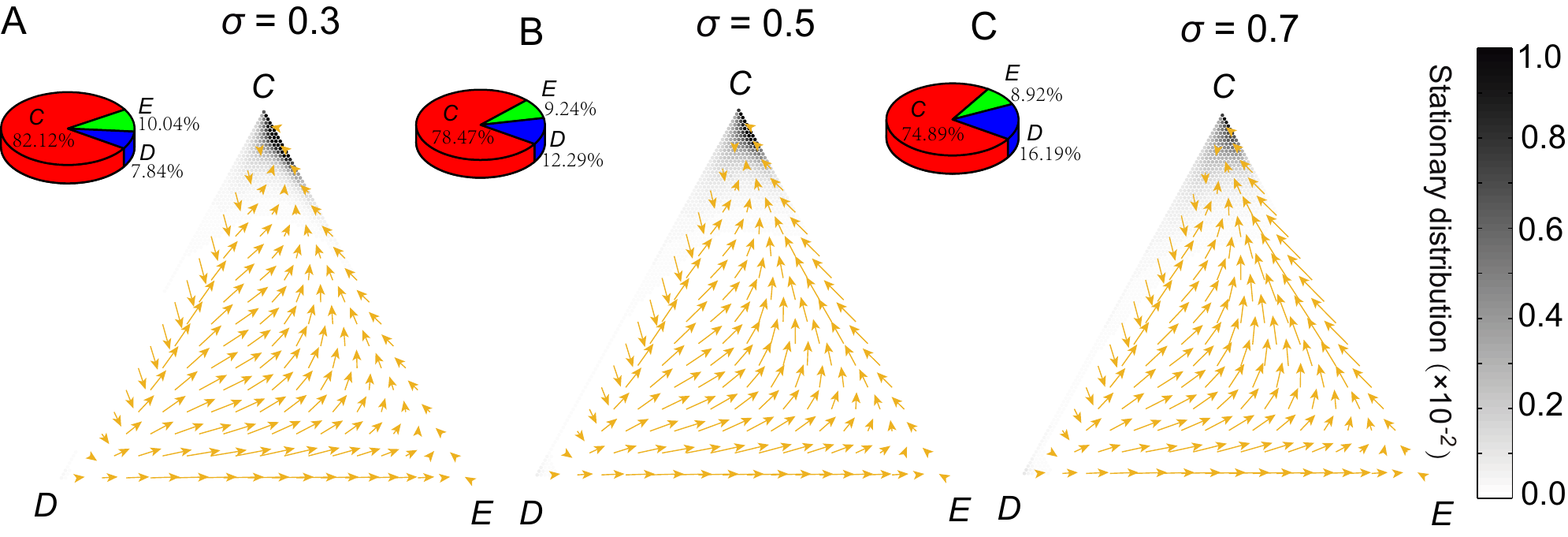}
\begin{flushleft}
\textbf{Supplementary Figure 8: Evolutionary dynamics of cooperators, defectors, and excluders in a finite population for three different values of monitoring cost $\sigma$.} In each $3D$ pie chart, the average levels of these three strategies are present. The increase of $\sigma$ weakens the advantage of excluders over defectors, while the population still spends most time in configurations where cooperators are prevalent. Here, $\sigma=0.3$ in panel A, $\sigma=0.5$ in panel B, and $\sigma=0.7$ in panel C. Other parameters values are $Z=100$, $N=5$, $F=3$, $c=1$, $c_{E}=0.4$, $\varsigma=2$, $w=0.8$, $\beta=2$, and $\mu=10^{-2}$.
\end{flushleft}
\label{figs7}
\end{figure*}

In Fig.~S8, we illustrate how different monitoring costs affect the stochastic dynamics of these three strategies. For small monitoring cost $\sigma=0.3$, the population spends most of the time in configurations near the state of all cooperation (Fig.~S8A). The increase of monitoring cost weakens the evolutionary advantages of excluders over defectors, and thus reduces the average level of cooperation (Fig.~S8B and C). However, it is worth emphasizing that the population still spends most of time in configurations of widespread cooperation (see 3$D$ pie chars).

In our main text, we investigate the evolutionary dynamics of cooperators, defectors, and excluders for a specific mutation rate $\mu$ with three different values of exclusion round $\varsigma$. However, it is necessary to explore the effects of different mutation rates on evolutionary outcomes. To this end, we now study the stationary distributions of the system and the corresponding gradient of selection for different mutation rates. The results are shown in Fig.~S9. Figure~S9A shows that, for smaller $\mu=10^{-3}$, the population spends most of the time in configurations which are close to the three corners of the simplex. As $\mu$ increases, the population explores configurations in which there are almost cooperators and excluders, as show in Fig.~S9B.

\begin{figure*}
\centering
\includegraphics[width=\textwidth]{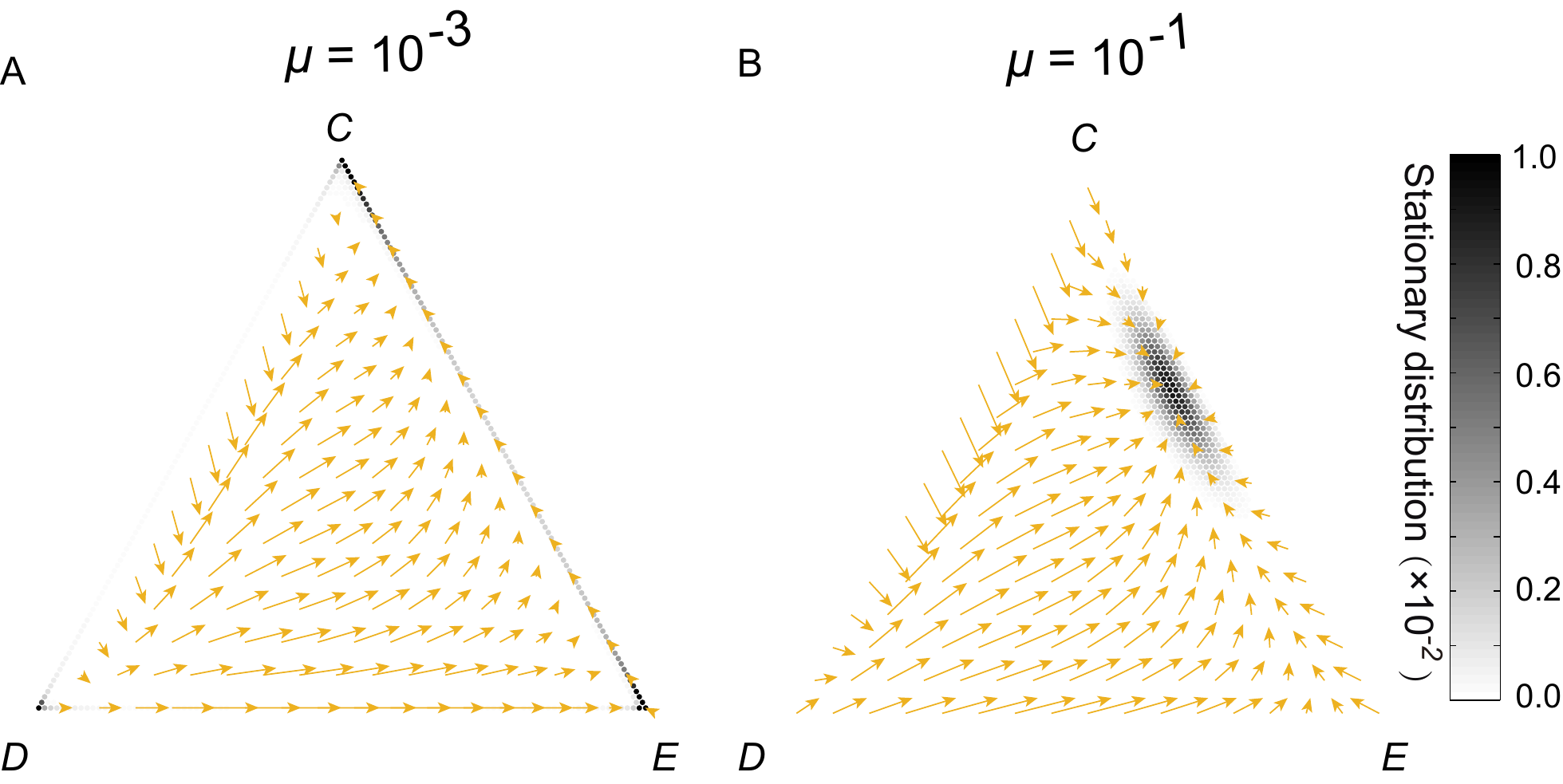}
\begin{flushleft}
\textbf{Supplementary Figure 9: Evolutionary dynamics of cooperators, defectors, and excluders in a finite population for different mutation rates $\mu$.} When mutation rate is small, the population spends more time in configurations corresponding to three vertices of the simplex. With the increase of mutation rate, the population explores configurations deviating gradually from the vertexes. Here, $\mu=10^{-3}$ in panel A and $\mu=10^{-1}$ in panel B. Other parameters values are $Z=100, N=5$, $F=3$, $c=1$, $c_{E}=0.4, \sigma=0.1$, $w=0.8, \beta=2$, and $\varsigma=2$.
\end{flushleft}
\label{figS8}
\end{figure*}

\noindent\subsection{\textbf{Sufficiently small mutation rates}}

Then, we assume that the mutation rate $\mu$ is sufficiently small.
It implies that the population will not contain more than two different strategies simultaneously, and the population can evolve into a homogeneous state where all individuals adopt the same strategy because the time between two mutation events is long enough. Thus the evolutionary dynamics can be determined by means of an embedded Markov chain in which the states correspond to the different homogeneous states of the population. If the number of available strategies
is denoted by $d$, then the state transition matrix which contains the different transition probabilities for the population to move from one state to another state is given by
\begin{eqnarray*}
\textbf{A}=[a_{UV}]_{d\times d},
\end{eqnarray*}
where $a_{UV}$ is the probability that the system switches from the $U$ state (where all players adopt strategy $U$) to the $V$ state (where all players adopt strategy $V$) after the emergence of a single mutation. Here $a_{UV}=\frac{\rho_{UV}}{d-1}$ if $U\neq V$ and $a_{UV}=1-\sum_{U\neq V}\frac{\rho_{UV}}{d-1}$ otherwise, where $\rho_{UV}$ is the fixation probability that a single individual with strategy $V$ takes over a resident population of players with strategy $U$. The stationary distribution of the system can be calculated from the average fraction of time that population spends in each of the homogeneous states. Technically, it is given by the normalized left eigenvector of the eigenvalue $1$ of the transition matrix $\textbf{A}$.

The fixation probability $\rho_{UV}$ can be calculated as follows. We assume that in a finite population with size $Z$ there are $i_{U}$ players using strategy $U$ and $i_{V}=Z-i_{U}$ players using strategy $V$. Then the probability that the number of players who adopt strategy $V$ increases/decreases by one is
\begin{eqnarray}
T^{\pm}(i_{V})=\frac{i_{V}}{Z}\frac{i_{U}}{Z}\frac{1}{1+\exp{[\mp
\beta(f_{VU}-f_{UV})]}},
\end{eqnarray}
where $f_{VU}$ is the average payoff of players with strategy $V$ against players with strategy $U$. Correspondingly, the fixation probability $\rho_{UV}$ can be written as
\begin{align}
\rho_{UV}&=\frac{1}{1+\sum_{q=1}^{Z-1}\prod_{i_{V}=1}^{q}\frac{T^{-}(i_{V})}{T^{+}(i_{V})}}\nonumber\\
&=\frac{1}{1+\sum_{q=1}^{Z-1}\exp{[\beta\sum_{i_{V}=1}^{q}(f_{UV}-f_{VU})]}}.
\end{align}

In the following, we present the payoff expressions for all the possible pairs in a finite population. In particular, when the population is consisted of pure cooperators and defectors, then the average payoffs of $C$ and $D$ players can be respectively written as
\begin{eqnarray*}
f_{CD}&=&\sum_{N_{C}=0}^{N-1}\frac{\binom{i_{C}-1}{N_{C}}\binom{Z-i_{C}}{N-N_{C}-1}}{\binom{Z-1}{N-1}}[\frac{F\langle r\rangle(N_{C}+1)c}{N}-\langle r\rangle c]\nonumber\\
&=&\frac{F\langle r\rangle c}{N}[\frac{(N-1)(i_{C}-1)}{Z-1}+1]-\langle r\rangle c,\\
f_{DC}&=&\sum_{N_{C}=0}^{N-1}\frac{\binom{i_{C}}{N_{C}}\binom{Z-i_{C}-1}{N-N_{C}-1}}{\binom{Z-1}{N-1}}\frac{FN_{C}\langle r\rangle c}{N}\nonumber\\
&=&\frac{F\langle r\rangle c}{N}\frac{(N-1)i_{C}}{Z-1},
\end{eqnarray*}
where $N_{C}$ denotes the number of cooperators in the group. The hypergeometric distribution $\frac{\binom{i_{C}-1}{N_{C}}\binom{Z-i_{C}}{N-N_{C}-1}}{\binom{Z-1}{N-1}}$ denotes the probability to find $N_{C}$ cooperators and $N-N_{C}-1$ defectors in the group.

Similarity, when cooperators compete against excluders, the average payoffs of these two strategies can be respectively written as
\begin{eqnarray*}
f_{CE}&=&F\langle r\rangle c-\langle r\rangle c,\\
f_{EC}&=&F\langle r\rangle c-\langle r\rangle c-\sigma.
\end{eqnarray*}
When excluders compete against defectors, the average payoffs of $E$ and $D$ can be respectively written as
\begin{eqnarray*}
f_{ED}&=&\frac{F(\varsigma-1)c}{N}[\frac{(N-1)(i_{E}-1)}{Z-1}+1]+Fc(\langle r\rangle-\varsigma+1)-\langle r\rangle c-c_{E}\frac{(N-1)(Z-i_{E})}{Z-1}-\sigma,\\
f_{DE}&=&\frac{F(\varsigma-1)c}{N}\frac{(N-1)i_{E}}{Z-1}.
\end{eqnarray*}

In finite populations, it has been addressed that the intensity of selection plays a critical role and can significantly influence the frequencies of competing strategies~\cite{sigmund2010social}. In the following, we calculate the average frequencies of $C, D$, and $E$ both for weak and strong selection in finite well-mixed populations.

\noindent\subsection{\textbf{Weak selection}}

In the limit case of weak selection ($\beta\rightarrow 0$), the linear approximation of the stationary distribution according to previous work \citep{sui2018rationality} is given as
\begin{eqnarray}
\Pi(\beta)=\frac{1}{d}+\frac{1}{d^{2}}
\left(
  \begin{array}{ccc}
    \sum_{s=1}^{d}\big(\sum_{i_{l}=1}^{Z-1}(f_{1s}(i_{l})-f_{s1}(i_{l}))\big) \\
    \sum_{s=1}^{d}\big(\sum_{i_{l}=1}^{Z-1}(f_{2s}(i_{l})-f_{s2}(i_{l}))\big)\\
    \sum_{s=1}^{d}\big(\sum_{i_{l}=1}^{Z-1}(f_{3s}(i_{l})-f_{s3}(i_{l}))\big)\\
    \vdots\\
    \sum_{s=1}^{d}\big(\sum_{i_{l}=1}^{Z-1}(f_{ds}(i_{l})-f_{sd}(i_{l}))\big)\\
  \end{array}
\right)\beta+\mathcal{O}(\beta),
\end{eqnarray}
where $\mathcal{O}(\beta)$ is the high-order infinitesimal of $\beta$. In combination with the above payoff expressions, we respectively show the stationary distributions of cooperators, defectors, and excluders for weak selection as
\begin{eqnarray*}
\Pi_{C}(\beta)&=&\frac{1}{3}+\frac{1}{9}[\sum_{i_{C}=1}^{Z-1}(f_{CD}-f_{DC})+\sum_{i_{C}=1}^{Z-1}(f_{CE}-f_{EC})]\beta+\mathcal{O}(\beta)\nonumber\\
&=&\frac{1}{3}+\frac{1}{9}(Z-1)[\frac{F\langle r\rangle c}{N}-\frac{F\langle r\rangle c(N-1)}{N(Z-1)}-\langle r\rangle c+\sigma]\beta+\mathcal{O}(\beta),\\
\Pi_{D}(\beta)&=&\frac{1}{3}+\frac{1}{9}[\sum_{i_{D}=1}^{Z-1}(f_{DC}-f_{CD})+\sum_{i_{D}=1}^{Z-1}(f_{DE}-f_{ED})]\beta+\mathcal{O}(\beta)\nonumber\\
&=&\frac{1}{3}+\frac{1}{9}(Z-1)[\frac{F\langle r\rangle c(N-1)}{N(Z-1)}+2\langle r\rangle c+\frac{Fc(N-1)(\varsigma-1)}{N(Z-1)}+\frac{Zc_{E}(N-1)}{2(Z-1)}+\sigma\nonumber\\
&-&\frac{F\langle r\rangle c}{N}-\frac{Fc(\varsigma-1)}{N}-Fc(\langle r\rangle-\varsigma+1)]\beta+\mathcal{O}(\beta),\\
\Pi_{E}(\beta)&=&\frac{1}{3}+\frac{1}{9}[\sum_{i_{E}=1}^{Z-1}(f_{ED}-f_{DE})+\sum_{i_{E}=1}^{Z-1}(f_{EC}-f_{CE})]\beta+\mathcal{O}(\beta)\nonumber\\
&=&\frac{1}{3}+\frac{1}{9}(Z-1)[-2\sigma+\frac{Fc(\varsigma-1)}{N}-\frac{Fc(\varsigma-1)(N-1)}{N(Z-1)}+Fc(\langle r\rangle-\varsigma+1)\nonumber\\
&-&\langle r\rangle c-\frac{Zc_{E}(N-1)}{2(Z-1)}]\beta+\mathcal{O}(\beta).
\end{eqnarray*}

Then the rank of the three strategies under weak selection depends on the relationship of the following three formulas: $\frac{F\langle r\rangle c}{N}-\frac{F\langle r\rangle c(N-1)}{N(Z-1)}-\langle r\rangle c+\sigma, \frac{F\langle r\rangle c(N-1)}{N(Z-1)}+2\langle r\rangle c+\frac{Fc(N-1)(\varsigma-1)}{N(Z-1)}+\frac{Zc_{E}(N-1)}{2(Z-1)}+\sigma-\frac{F\langle r\rangle c}{N}-\frac{Fc(\varsigma-1)}{N}-Fc(\langle r\rangle-\varsigma+1)$, and $-2\sigma+\frac{Fc(\varsigma-1)}{N}-\frac{Fc(\varsigma-1)(N-1)}{N(Z-1)}+Fc(\langle r\rangle-\varsigma+1)-\langle r\rangle c-\frac{Zc_{E}(N-1)}{2(Z-1)}$. In the following, we present two examples to illustrate the rank of three strategies.

\noindent\textbf{Example 1:} Excluders can do better than cooperators, and cooperators have evolutionary advantage over defectors.

When $\frac{F\langle r\rangle c}{N}-\frac{F\langle r\rangle c(N-1)}{N(Z-1)}-\langle r\rangle c+\sigma<-2\sigma+\frac{Fc(\varsigma-1)}{N}-\frac{Fc(\varsigma-1)(N-1)}{N(Z-1)}+Fc(\langle r\rangle-\varsigma+1)-\langle r\rangle c-\frac{Zc_{E}(N-1)}{2(Z-1)}$, excluders can have evolutionary advantage over cooperators. By simplifying, we can get
$$\varsigma<\frac{-3\sigma+F\langle r\rangle c-Zc_{E}(N-1)/(2Z-2)-F\langle r\rangle c(Z-N)/(NZ-N)}{Fc-Fc/N+Fc(N-1)/(NZ-N)}+1.$$
Similarly, when $\frac{F\langle r\rangle c(N-1)}{N(Z-1)}+2\langle r\rangle c+\frac{Fc(N-1)(\varsigma-1)}{N(Z-1)}+\frac{Zc_{E}(N-1)}{2(Z-1)}+\sigma-\frac{F\langle r\rangle c}{N}-\frac{Fc(\varsigma-1)}{N}-Fc(\langle r\rangle-\varsigma+1)<\frac{F\langle r\rangle c}{N}-\frac{F\langle r\rangle c(N-1)}{N(Z-1)}-\langle r\rangle c+\sigma$, i.e.,
$$\varsigma<\frac{F\langle r\rangle c-3\langle r\rangle c-Zc_{E}(N-1)/(2Z-2)+2F\langle r\rangle c(Z-N)/(NZ-N)}{Fc-Fc/N+Fc(N-1)/(NZ-N)}+1,$$
cooperator can do better than defector. Therefore, we can give the condition in which $E$ can do better than $C$, and $C$ has evolutionary advantage over $D$:
\begin{eqnarray*}
\varsigma<\min\{\frac{-3\sigma+F\langle r\rangle c-Zc_{E}(N-1)/(2Z-2)-F\langle r\rangle c(Z-N)/(NZ-N)}{Fc-Fc/N+Fc(N-1)/(NZ-N)}+1, \\
\frac{F\langle r\rangle c-3\langle r\rangle c-Zc_{E}(N-1)/(2Z-2)+2F\langle r\rangle c(Z-N)/(NZ-N)}{Fc-Fc/N+Fc(N-1)/(NZ-N)}+1\}.
\end{eqnarray*}
For large population ($Z\gg N$), this condition reduces to
\begin{eqnarray*}
\varsigma<\min\{\frac{-3\sigma+F\langle r\rangle c-c_{E}(N-1)/2-F\langle r\rangle c/N}{Fc-Fc/N}+1, \\\frac{F\langle r\rangle c-3\langle r\rangle c-c_{E}(N-1)/2+2F\langle r\rangle c/N}{Fc-Fc/N}+1\}.
\end{eqnarray*}

\begin{figure*}
\centering
\includegraphics[width=\textwidth]{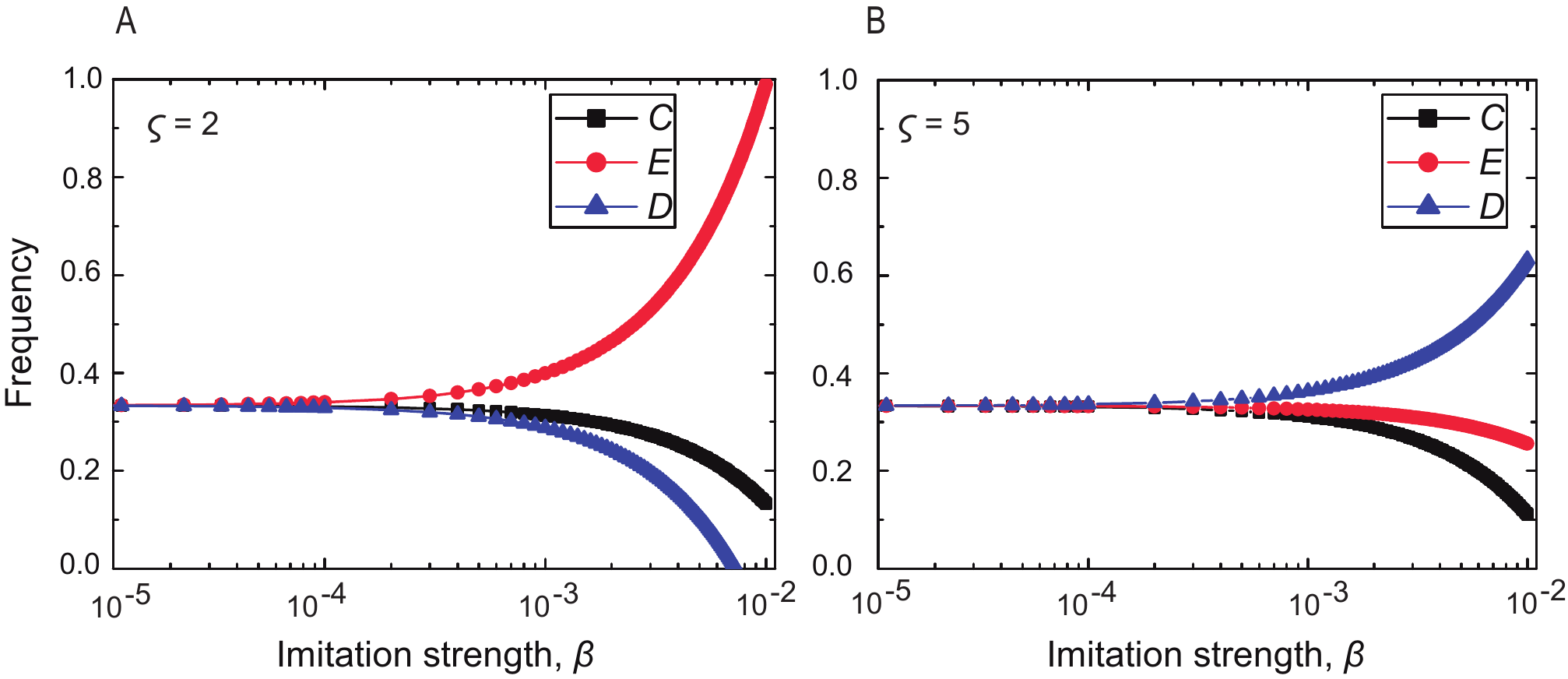}
\begin{flushleft}
\textbf{Supplementary Figure 10: Evolutionary dynamics of cooperators, defectors, and excluders in a finite population under weak selection for two different exclusion round $\varsigma$ values.} The rank among these three strategies can be determined by the exclusion round $\varsigma$. Panel A shows that exclusion prevails over cooperation and defection, while panel B shows that defection has evolutionary advantages over cooperation and exclusion. Here, $\varsigma=2$ in panel A and $\varsigma=5$ in panel B. Other parameters values are $Z=100, N=5$, $F=3$, $c=1$, $c_{E}=0.4, \sigma=0.1$, and $w=0.8$.
\end{flushleft}
\label{figS9}
\end{figure*}

Then we provide a numerical example to verify the theoretical result when the parameter values satisfy the condition given above. In Fig.~S10A we show the rank of cooperation, defection, and exclusion when the intensity of selection $\beta$ is weak. For sufficiently small $\beta$, the frequencies of the three competing strategies are identical due to the practically random updating process. With the increase of $\beta$, the rank of the three strategies changes gradually. As a result, excluders form the highest portion, and the second largest portion is formed by cooperators, while defectors can make up the smallest fraction. All these results are in agreement with the theoretical prediction.

\noindent\textbf{Example 2:} Defectors can do better than excluders and excluders have evolutionary advantage over cooperators.

When $-2\sigma+\frac{Fc(\varsigma-1)}{N}-\frac{Fc(\varsigma-1)(N-1)}{N(Z-1)}+Fc(\langle r\rangle-\varsigma+1)-\langle r\rangle c-\frac{Zc_{E}(N-1)}{2(Z-1)}<\frac{F\langle r\rangle c(N-1)}{N(Z-1)}+2\langle r\rangle c+\frac{Fc(N-1)(\varsigma-1)}{N(Z-1)}+\frac{Zc_{E}(N-1)}{2(Z-1)}+\sigma-\frac{F\langle r\rangle c}{N}-\frac{Fc(\varsigma-1)}{N}-Fc(\langle r\rangle-\varsigma+1)$, defectors can have evolutionary advantage over excluders. By simplifying, we can get
\begin{eqnarray*}
\varsigma>\frac{-3\sigma+2F\langle r\rangle c-Zc_{E}(N-1)/(Z-1)-3\langle r\rangle c+F\langle r\rangle c(Z-N)/(NZ-N)}{2[Fc-Fc/N+Fc(N-1)/(NZ-N)]}+1.
\end{eqnarray*}
As we stated above, when $\varsigma<\frac{-3\sigma+F\langle r\rangle c-Zc_{E}(N-1)/(2Z-2)-F\langle r\rangle c(Z-N)/(NZ-N)}{Fc-Fc/N+Fc(N-1)/(NZ-N)}+1$, excluder can do better than cooperator. Therefore, we can give the condition in which $D$ can do better than $E$, and $E$ has evolutionary advantage over $C$:
\begin{eqnarray*}
\frac{-3\sigma+2F\langle r\rangle c-Zc_{E}(N-1)/(Z-1)-3\langle r\rangle c+F\langle r\rangle c(Z-N)/(NZ-N)}{2[Fc-Fc/N+Fc(N-1)/(NZ-N)]}+1<\varsigma\\
<\frac{-3\sigma+F\langle r\rangle c-Zc_{E}(N-1)/(2Z-2)-F\langle r\rangle c(Z-N)/(NZ-N)}{Fc-Fc/N+Fc(N-1)/(NZ-N)}+1.
\end{eqnarray*}
For large population ($Z\gg N$), this condition reduces to
\begin{eqnarray*}
\frac{-3\sigma+2F\langle r\rangle c-c_{E}(N-1)-3\langle r\rangle c+F\langle r\rangle c/N}{2(Fc-Fc/N)}+1<\varsigma\\
<\frac{-3\sigma+F\langle r\rangle c-c_{E}(N-1)/2-F\langle r\rangle c/N}{Fc-Fc/N}+1.
\end{eqnarray*}
Next we provide a numerical example to confirm that the frequency of defectors is higher than that of excluders, and the frequency of excluders is higher than that of cooperators when the above condition is satisfied. As shown in Fig.S10B, when the intensity of selection is weak ($\beta<10^{-4}$),  the frequencies of cooperators, defectors, and excluders are identical. With the increase of $\beta$, the advantage of defectors increases gradually. Consequently, the frequency of defectors increases and the frequencies of cooperators and excluders both decrease. Importantly, we can find that the reduction of cooperators is larger than that of excluders.

\noindent\subsection{\textbf{Strong selection}}

In the limit of strong selection (i.e. $\beta\rightarrow \infty$), the fixation probability is simplified and its value is closely related to the average payoff difference. Here the fixation probability $\rho_{UV}=1$ when $f_{UV}<f_{VU}$, and $\rho_{UV}=\frac{1}{Z}$ when $f_{UV}=f_{VU}$, otherwise $\rho_{UV}=0$. Then, the transition probability from $AllU$ to $AllV$ can be given as $a_{UV}=\frac{1}{d-1}$, $\frac{1}{Z(d-1)}$, or $0$, respectively.

Next, we present the elements in the transition matrix $\textbf{A}$. For large populations, a single cooperator always has a smaller payoff than defectors. Thus a single cooperator cannot invade a resident population of defectors, which leads to $\rho_{DC}=0$. Then we have $a_{DC}=0$. Conversely, a single defector has a higher payoff than a resident population of cooperators. In this case, $\rho_{CD}=1$, which leads to $a_{CD}=\frac{1}{2}$.
Besides, it is certain that a single defector can invade a resident population of excluders when $\varsigma>\frac{N(Z-1)(F\langle r\rangle c-\langle r\rangle c-\sigma)-c_{E}N(N-1)}{Fc(N-1)Z}+1$. While when $\varsigma<\frac{N(Z-1)(F\langle r\rangle c-\langle r\rangle c-\sigma)-c_{E}N(N-1)(Z-1)}{Fc(N-1)Z}+1$,
a single excluder has a larger payoff than a resident population of defectors. Thus we summarize that $a_{ED}=\frac{1}{2}$ and $a_{DE}=0$ when $\varsigma>\frac{N(Z-1)(F\langle r\rangle c-\langle r\rangle c-\sigma)-c_{E}N(N-1)}{Fc(N-1)Z}+1$. While when $\varsigma<\frac{N(Z-1)(F\langle r\rangle c-\langle r\rangle c-\sigma)-c_{E}N(N-1)(Z-1)}{Fc(N-1)Z}+1$, we have $a_{ED}=0$ and $a_{DE}=\frac{1}{2}$. Particularly, when $\frac{N(Z-1)(F\langle r\rangle c-\langle r\rangle c-\sigma)-c_{E}N(N-1)(Z-1)}{Fc(N-1)Z}+1<\varsigma<\frac{N(Z-1)(F\langle r\rangle c-\langle r\rangle c-\sigma)-c_{E}N(N-1)}{Fc(N-1)Z}+1$, one defector cannot invade a resident population of excluders and meanwhile one excluder also cannot invade a resident population of defectors, thus we have $a_{ED}=0$ and $a_{DE}=0$.

Based on the above theoretical analysis, we summarize the transitions matrix among $C, D$, and $E$ strategies as follows.

\textbf{Case 1}: For $\varsigma<\frac{N(Z-1)(F\langle r\rangle c-\langle r\rangle c-\sigma)-c_{E}N(N-1)(Z-1)}{Fc(N-1)Z}+1$, in the case of strong selection, the transition matrix among cooperators ($C$), excluders ($E$), and defectors ($D$) is
\begin{equation}
\bordermatrix{%
          && C              && E                &&  D           \cr
C    && \frac{1}{2}         && 0                &&\frac{1}{2}   \cr
E    && \frac{1}{2}         && \frac{1}{2}      &&0             \cr
D    && 0                   && \frac{1}{2}      &&\frac{1}{2}
}.
\end{equation}
Accordingly, the stationary distribution is $(\frac{1}{3}, \frac{1}{3}, \frac{1}{3})$.

\textbf{Case 2}: For $\frac{N(Z-1)(F\langle r\rangle c-\langle r\rangle c-\sigma)-c_{E}N(N-1)(Z-1)}{Fc(N-1)Z}+1<\varsigma<\frac{N(Z-1)(F\langle r\rangle c-\langle r\rangle c-\sigma)-c_{E}N(N-1)}{Fc(N-1)Z}+1$, in the case of strong selection, the transition matrix among cooperators ($C$), excluders ($E$), and defectors ($D$) is
\begin{equation}
\bordermatrix{%
          && C              && E                &&  D           \cr
C    && \frac{1}{2}         && 0                &&\frac{1}{2}   \cr
E    && \frac{1}{2}         && \frac{1}{2}      &&0             \cr
D    && 0                   && 0                &&1
}.
\end{equation}
Accordingly, the stationary distribution is $(0, 0, 1)$.

\textbf{Case 3}: For $\varsigma>\frac{N(Z-1)(F\langle r\rangle c-\langle r\rangle c-\sigma)-c_{E}N(N-1)}{Fc(N-1)Z}+1$, in the case of strong selection, the transition matrix among cooperators ($C$), excluders ($E$), and defectors ($D$) is
\begin{equation}
\bordermatrix{%
          && C              && E                &&  D           \cr
C    && \frac{1}{2}         && 0                &&\frac{1}{2}   \cr
E    && \frac{1}{2}         && 0                &&\frac{1}{2}   \cr
D    && 0                   && 0                &&1
}.
\end{equation}
Accordingly, the stationary distribution is $(0, 0, 1)$.

\begin{figure*}
\centering
\includegraphics[width=\textwidth]{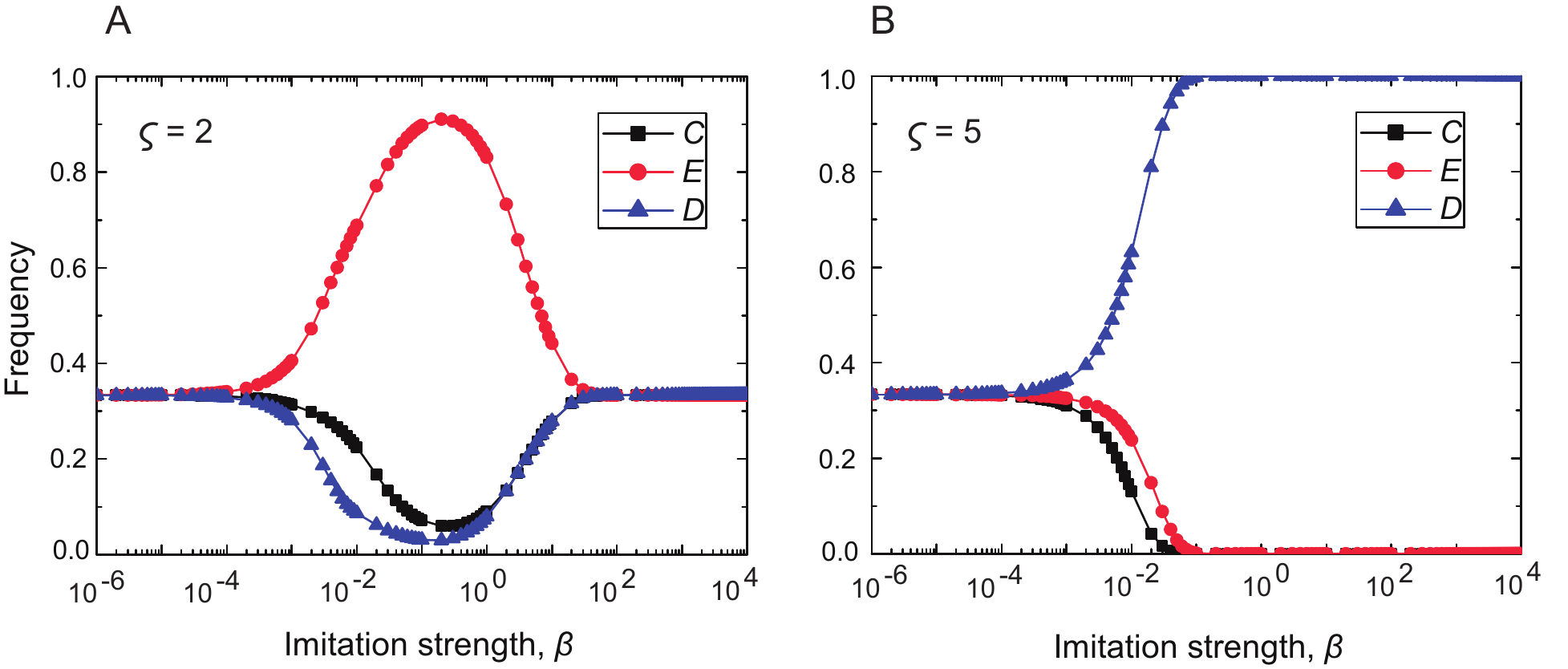}
\begin{flushleft}
\textbf{Supplementary Figure 11: Evolutionary dynamics of cooperators, defectors, and excluders in a finite population for two different values of exclusion round $\varsigma$.} The rank among these three strategies can be determined by the intensity of selection $\beta$ and the exclusion round $\varsigma$. Panel A shows that the frequency of each strategy tends to be $1/3$ for strong selection, while panel B shows that defection prevails over cooperation and exclusion for strong selection. Here, $\varsigma=2$ in panel A and $\varsigma=5$ in panel B. Other parameters values are $Z=100, N=5$, $F=3$, $c=1$, $c_{E}=0.4, \sigma=0.1$, and $w=0.8$.
\end{flushleft}
\label{figs10}
\end{figure*}

We then provide numerical examples to verify our theoretical analysis. Here we only provide numerical results for Case~1 and Case~2 since the result for Case~3 is the same to that for Case~2 for strong selection. In Fig.~S11, we illustrate the rank of $C, E$, and $D$ for two different values of exclusion round $\varsigma$. For small $\varsigma$, when the intensity of selection is weak, the numerical result is in agreement with the theoretical prediction result (Fig.~S10A and Fig.~S11A), that is, excluders have evolutionary advantages over cooperators and defectors. However, the evolutionary advantage of excluders will be weakened by the increase of $\beta$. Particularly, for strong selection the frequency curves of the three strategies will overlap again. For large $\varsigma$, defectors prevail over excluders and cooperators for strong selection (Fig.~S11B).


\vbox{}

\end{document}